\documentclass[compress]{article}

\usepackage{arxiv}
\usepackage[utf8]{inputenc} 
\usepackage[T1]{fontenc}    
\usepackage{hyperref}       
\usepackage{url}            
\usepackage{booktabs}       
\usepackage{amsfonts}       
\usepackage{nicefrac}       
\usepackage{microtype}      
\usepackage{lipsum}
\usepackage{amsmath}        
\usepackage{graphicx}       
\graphicspath{ {images/} }
\usepackage{subcaption}
\usepackage{float}
\usepackage{multirow}       
\usepackage{array}
\newcolumntype{P}[1]{>{\centering\arraybackslash}p{#1}}
\newcolumntype{M}[1]{>{\centering\arraybackslash}m{#1}}

\title{Compressive-Sensing Data Reconstruction for Structural Health Monitoring: A Machine-Learning Approach}

\author{
  Yuequan Bao\thanks{Correspondence to: Dr. Yuequan Bao (Professor), School of Civil Engineering, Harbin Institute of Technology, Harbin, China. Artificial Intelligence Lab, Harbin Institute of Technology, Harbin, China. E-mail: baoyuequan@hit.edu.cn} \\
  School of Civil Engineering\\
  Harbin Institute of Technology\\
  Harbin, China 150090 \\
  \texttt{baoyuequan@hit.edu.cn} \\
  \And
  Zhiyi Tang \\
  School of Civil Engineering\\
  Harbin Institute of Technology\\
  Harbin, China 150090 \\
  \texttt{tang@stu.hit.edu.cn} \\
  \And
  Hui Li \\
  School of Civil Engineering\\
  Harbin Institute of Technology\\
  Harbin, China 150090 \\
  \texttt{lihui@hit.edu.cn} \\
}

\begin{document}

\setlength\abovedisplayskip{0pt}
\setlength\belowdisplayskip{0pt}

\maketitle

\begin{abstract}
Compressive sensing (CS) has been studied and applied in structural health monitoring for data acquisition and reconstruction, wireless data transmission, structural modal identification, and spare damage identification. The key issue in CS is finding the optimal solution for sparse optimization. In the past several years, many algorithms have been proposed in the field of applied mathematics. In this paper, we propose a machine-learning-based approach to solve the CS data-reconstruction problem. By treating a computation process as a data flow, the solving process of CS-based data reconstruction is formalized into a standard supervised-learning task. The prior knowledge, i.e., the basis matrix and the CS-sampled signals, are used as the input and the target of the network; the basis coefficient matrix is embedded as the parameters of a certain layer; and the objective function of conventional compressive sensing is set as the loss function of the network. Regularized by $l_{1}$-norm, these basis coefficients are optimized to reduce the error between the original CS-sampled signals and the masked reconstructed signals with a common optimization algorithm. In addition, the proposed network is able to handle complex bases, such as a Fourier basis. Benefiting from the nature of a multi-neuron layer, multiple signal channels can be reconstructed simultaneously. Meanwhile, the disassembled use of a large-scale basis makes the method memory-efficient. A numerical example of multiple sinusoidal waves and an example of field-test wireless data from a suspension bridge are carried out to illustrate the data-reconstruction ability of the proposed approach. The results show that high reconstruction accuracy can be obtained by the machine learning-based approach. In addition, the parameters of the network have clear meanings; the inference of the mapping between input and output is fully transparent, making the CS data reconstruction neural network interpretable.
\end{abstract}

\vskip 0.5cm

\keywords{Structural health monitoring \and Compressive sensing \and Model-driven machine learning \and interpretable machine learning \and Sparse data reconstruction}

\vskip 0.5cm

\section{Introduction}
Structural health monitoring (SHM) is a technology used to maintain the safety of structures that has been widely used in aerospace, civil, and mechanical engineering~\citep{RN67,RN68,RN69,RN70,RN71,RN72,RN73,RN74}. Generally, a SHM system consists of various sensors, a data-acquisition and -transmission system, data analysis and modeling, structural health diagnosis (including data processing, data mining, damage detection, model updating, safety evaluation, and reliability analysis), alarming devices, a visualization user interface, and software as well as an operating system~\citep{RN73}.

Data acquisition is the fundamental part in a SHM system. Traditional data acquisition must obey the Nyquist-Shannon sampling theorem, which requires a signal to be sampled at least two times at the highest frequency in the signal. Thus, the amount of data collected by the SHM system will be huge. Compressive sensing (CS), also known as compressive sampling as proposed by Donoho and Candes, provides a new sampling theory for signals with sparse features in a certain domain~\citep{RN13,RN14}. The signal is randomly collected, the size of which is much smaller than that recorded following the Nyquist-Shannon sampling theorem. Then, the original signal then can be exactly reconstructed with sparse optimization algorithms~\citep{RN13,RN14}.

In CS theory, the raw signal $y \in \mathbb{R}^n$  can be sampled using a linear measurement,

\vskip -0.3cm
\begin{equation} \label{eq:1}
y_s = \Phi y + e,
\end{equation}

\noindent where $\Phi$ denotes a measurement matrix or sampling operator in an $m \times n$ matrix, $y_s$ is the sampled signal vector, and e is the measurement noise.

As $\Phi$ is an $m \times n$ matrix with $m \ll n$, the problem of reconstructing the signal $x$ is ill-posed. However, CS theory proved that if the signal is sparse (i.e., the signal has a sparse representation in some basis $\Psi$ or a redundant dictionary $D$, $y=\Psi x$ or $y=D x$) and the $\Phi$ satisfies the restrictive isometry property (RIP), then the coefficient can be reconstructed by the l1 optimization problem:

\vskip -0.3cm
\begin{equation} \label{eq:2}
y=\Phi \Psi x + e = \Theta x + e,
\end{equation}

\vskip -0.5cm
\begin{equation} \label{eq:3}
\hat{x} = \min \Vert x \Vert_1 \text{  such that  } \Vert y - \Theta \hat{x} \Vert_2 \leq \varepsilon,
\end{equation}

\noindent where $\varepsilon$ is the bound on the level of the measurement error, $\Vert e \Vert_2 \leq \varepsilon$ .

Many algorithms have been proposed to solve the optimization problem of equation (\ref{eq:3}), such as basic pursuit (BP)~\citep{RN15}, gradient projection for sparse reconstruction (GPSR)~\citep{RN16}, l1 regularized least squares (l1-ls)~\citep{RN82}, fixed-point continuation (FPC)~\citep{RN17} and the FPC active set (FPC-AS)~\citep{RN44}, split Bregman~\citep{RN43}, the fast iterative shrinkage-thresholding algorithm~\citep{RN18}, sparse reconstruction by separable approximation (SpaRSA)~\citep{RN20}, NESTA (Nesterov’s algorithm)~\citep{RN45}, the Bayesian methods~\citep{RN22}, and the group least absolute shrinkage and selection operator (LASSO)~\citep{RN47}.

CS has been widely used in many fields, including consumer camera imaging~\citep{RN80,RN81}, medical magnetic-resonance imaging~\citep{RN82}, remote sensing~\citep{RN83}, seismic exploration~\citep{RN85}, and communications, especially for wireless sensor networks (WSN)~\citep{RN86,RN87}. In SHM, the applications of CS theory have also been investigated in structural vibration data acquisition~\citep{RN25}, wireless data transmission and lost data recovery~\citep{RN24,RN28,RN29,RN31,RN27,RN23}, structural modal identification~\citep{RN32,RN66}, structural sparse damage identification~\citep{RN35,RN36,RN37,RN38,RN39,RN40}, and sparse heavy-vehicle-loads identification of long-span bridges~\citep{RN41}.

As mentioned above, the nature of CS is to solve an ill-posed inverse problem. By constructing an efficient objective function and regularization, the original data can be reconstructed by iterative optimization procedures. The recent advanced deep-learning techniques, as iterative methods, are naturally exploited to solve ill-posed inverse problems, especially in imaging, including image reconstruction~\citep{RN63,RN92,RN77}, super-resolution~\citep{RN78,RN91,RN90}, image denoising and inpainting~\citep{RN91,RN90}, and image colorization~\citep{RN89,RN88}. The state-of-the-art performance of the above-listed deep neural networks (DNNs) are mainly accomplished by a large-scale data-driven strategy; meanwhile, the inner workings of the DNNs continue to be difficult to understand and interpret. More recently, several works have been devoted to incorporate machine-learning techniques and domain prior knowledge~\citep{RN63,RN77,RN59}, which not only accomplished superior performance for inverse-problem solving, but also have interpretable inner mechanisms. To date, little effort has been made to develop the hybrid strategy outside the field of image processing and computer vision. In this paper, by treating a computation process as a data flow~\citep{RN58}, we formalize the CS problem into a standard supervised-learning task for time-series signal reconstruction. We design a fully transparent neural network embedded with   regularization. Unlike conventional black-box neural networks, the parameters of the network are interpretable with clear meaning; the inference of the mapping between input and output is fully transparent; and the large-scale basis matrix is distributedly utilized, which makes the method memory-efficient. In addition, the proposed network is able to handle complex bases, such as a Fourier basis. Benefiting from the multi-neuron layer, multiple channels of a signal can be reconstructed simultaneously.

The rest of the paper is organized as follows. The formalization of compressive sensing into a machine-learning task is explained in Section 2.1. Detailed architecture of the network and mathematics are given in Section 2.2. For method validation, a numerical example using synthetic signals is presented and discussed in Section 3.1, followed by another example using extensive field test wireless data in Section 3.2. Finally, discussion and conclusions are drawn in Section 4.

\section{Machine-learning approach for CS data reconstruction}

\subsection{Formalizing CS into a machine-learning task}

Suppose there are $K$ sensors (channels) implemented on a structure. Each channel of the sensors is sampled $N$ points uniformly in duration $T$, i.e., the sampling frequency is $f_s = N / T$. Collecting all the data, we obtain an $N \times K$ matrix $U$,

\begin{equation} \label{eq:4}
U = \begin{bmatrix}
u_{11} & \dots & u_{1K} \\
\vdots & u_{nk} & \vdots \\
u_{N1} & \dots & u_{NK}
\end{bmatrix}
, U \in \mathbb{R}^{N \times K},
\end{equation}

\noindent where $u_{nk}$ is the data measured by the $k$th sensor at time $t_n$.

Then, let $P$ denote a masking matrix to random sample the original signals $U$ via element-wise multiplication, in which all elements inside are either 0 or 1 in the Bernoulli distribution with sampling ratio $\gamma$ [equation (\ref{eq:5})]. The sampled signal matrix $Y$ is then $Y=P \odot U$, and the problem is how to calculate the original signal matrix $U$ from the given signal matrix $Y$:

\begin{equation} \label{eq:5}
f(p_{nk}) = \begin{cases}
\gamma & \text{for } p_{nk} = 1 \\
1 - \gamma & \text{for } p_{nk} = 0
\end{cases}
, p_{nk} \in P \in \mathbb{Z}^{N \times K}.
\end{equation}

As mentioned above, CS is an ill-posed inverse problem with a general representation like equation (\ref{eq:6}). When facing multiple signals simultaneously, which is commonly encountered, the CS problem of equation (\ref{eq:3}) can be represented based on “simultaneous sparsity”~\citep{RN61,RN62} or “joint sparsity”~\citep{RN60} in different areas as follows:

\begin{equation} \label{eq:6}
U = \Phi X + e, \Phi \in \mathbb{C}^{N \times N}, X \in \mathbb{C}^{N \times K}, e \in \mathbb{C}^{N \times K},
\end{equation}

\noindent where $X$ is the basis coefficient matrix of $N$ points and $K$ channels, $\Phi$ the basis matrix, and $e$ the error. The basis coefficient matrix $X$ is reconstructed by solving the following optimization problem~\citep{RN27}:

\begin{equation} \label{eq:7}
\min_{X \in \mathbb{C}^{N \times K}} \Vert \hat{X} \Vert_1 + \frac{\mu}{2} \Vert P \odot U - P \odot \Phi \hat{X} \Vert_2,
\end{equation}

\noindent where $\hat{X}$ is the possible solution of the Fourier coefficients matrix and $\mu$ the penalty weight. Once the optimal solution $X_{rec}$ is obtained, the reconstructed signal matrix is given by

\begin{equation} \label{eq:8}
U_{rec} = \Phi X_{rec}.
\end{equation}
\vskip 0.1cm

\begin{figure}[H]
\centering
\begin{subfigure}{0.4\textwidth}
\centering
\includegraphics[width=0.9\textwidth]{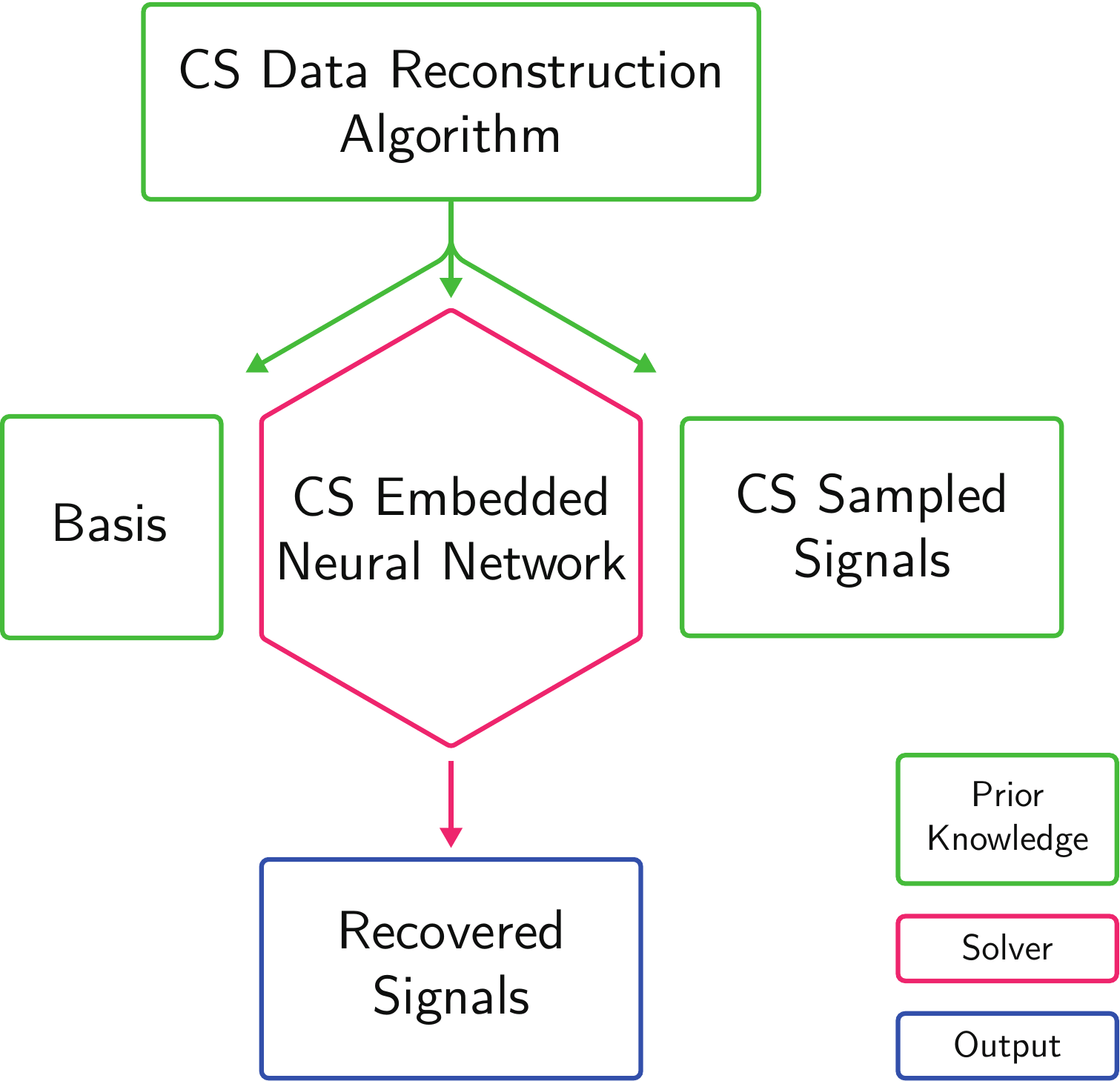} 
\caption{}
\label{fig:subim1}
\end{subfigure}
\hfill
\begin{subfigure}{0.5\textwidth}
\centering
\includegraphics[width=1.0\textwidth]{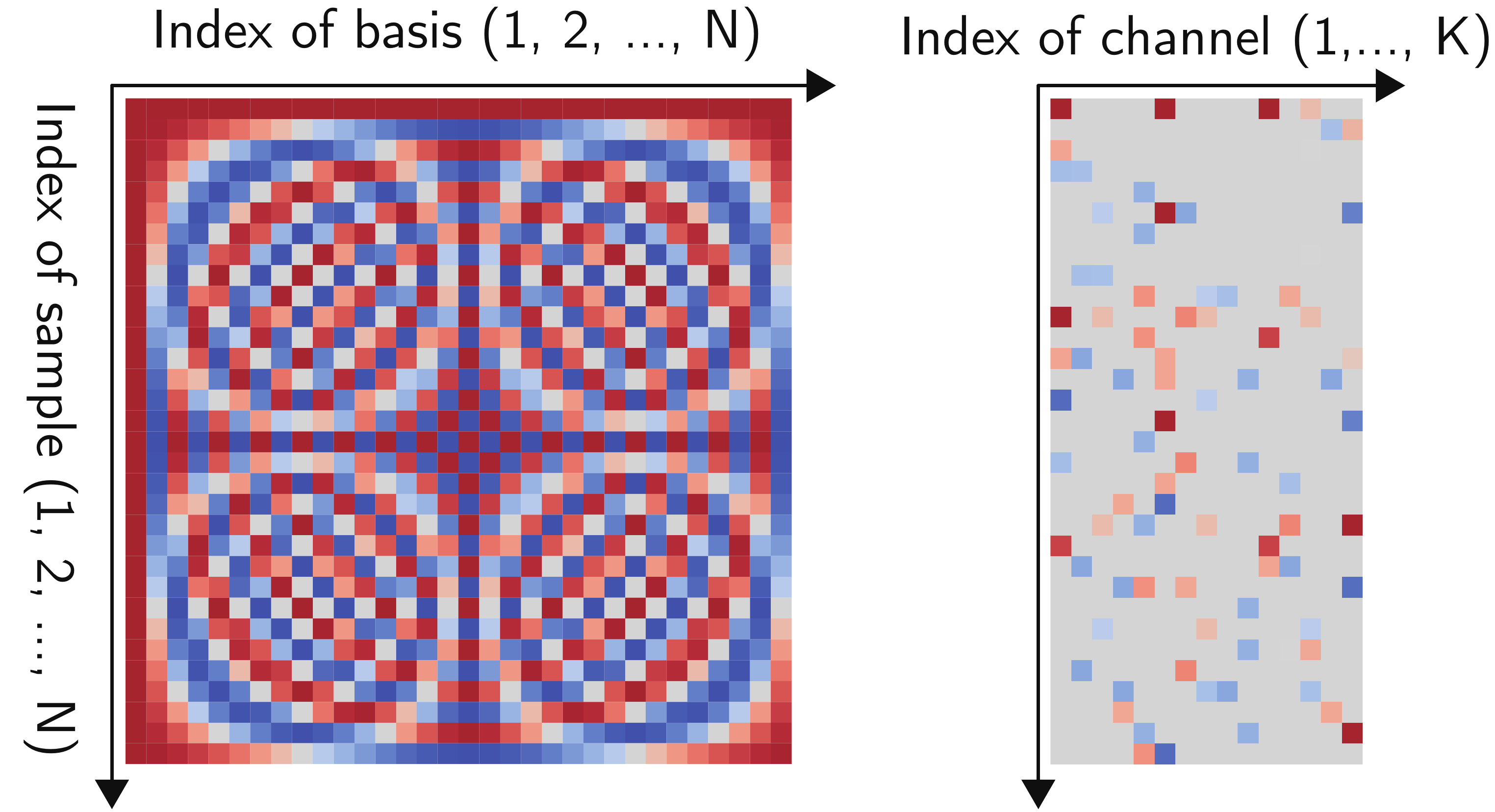}
\caption{}
\label{fig:subim2}
\end{subfigure}
\caption{(a) Mind map of formalizing CS problem into a supervised-learning task; (b) example of the input and output of the network for data reconstruction}
\label{fig:fig_1}
\end{figure}

It is shown that the form of equation (\ref{eq:7}) is equivalent to the loss function of a regression task with regularization of parameters in the context of machine learning. Figure 1(a) illustrates the mind map of formalizing CS into a supervised-learning task. One can consider the prior knowledge, i.e., the basis matrix and the CS-sampled signals as input and output, respectively; each row of them constitutes a pair of samples as shown in Figure 1(b). The basis coefficient matrix and masking matrix are embedded into a machine neural network as weights and neurons, respectively. Then, the computation process of conventional CS is constructed as the feed-forward of the network. Different from a regular black-box neural network, the proposed network for data reconstruction is prior-knowledge-embedded and interpretable for each layer. Mathematics and examples of the network are given in the following subsections.

\subsection{Construction of the novel CS-embedded neural network}

As shown in Figure 2, the proposed network consists of four modules: input, basis coefficients solving, masking, and output. In the first module, the basis matrix $\Phi$ acts as the input, and each row of the basis matrix is a training sample. For example, a basis matrix $\Phi$ of dimension $1024 \times 1024$ will generate 1024 samples, and the size of each sample is $1 \times 1024$. Note that the proposed network is able to handle complex bases, such as a Fourier basis. The real and imaginary parts of the basis are input into the network separately. If the basis matrix is real, then the imaginary-part inputs are all zero. Meanwhile, the disassembled use of the large-scale basis makes the method memory-efficient because there is no need to load the basis matrix into memory all at once.

\begin{figure}[H]
\centering
\includegraphics[width=0.8\textwidth]{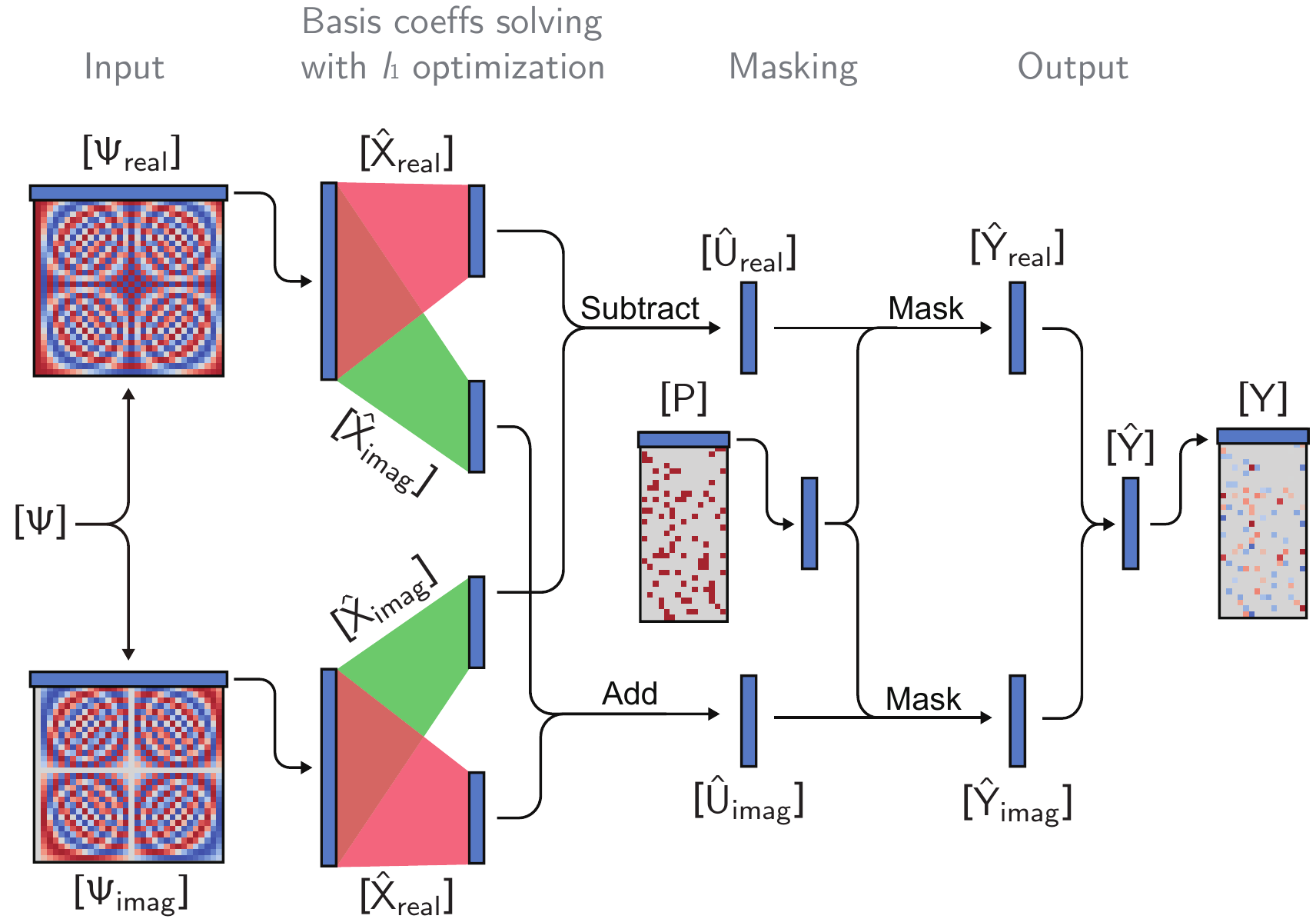}
\label{fig:fig_2}
\caption{Architecture of the proposed network for data reconstruction}
\end{figure}

\begin{equation} \label{eq:9}
\Psi_{real} = \begin{bmatrix}
\varPsi^{real}_{11} & \dots & \varPsi^{real}_{1N} \\
\vdots & \varPsi^{real}_{nn} & \vdots \\
\varPsi^{real}_{N1} & \dots & \varPsi^{real}_{NN}
\end{bmatrix}
, \Psi_{real} \in \mathbb{R}^{N \times N},
\end{equation}

\begin{equation} \label{eq:10}
\Psi_{imag} = \begin{bmatrix}
\varPsi^{imag}_{11} & \dots & \varPsi^{imag}_{1N} \\
\vdots & \varPsi^{imag}_{nn} & \vdots \\
\varPsi^{imag}_{N1} & \dots & \varPsi^{imag}_{NN}
\end{bmatrix}
, \Psi_{imag} \in \mathbb{R}^{N \times N},
\end{equation}

\begin{equation} \label{eq:11}
\Psi = \Psi_{real} + i \cdot \Psi_{imag}, \Psi \in \mathbb{C}^{N \times N}.
\end{equation}

Accordingly, the basis coefficient matrix $X$ is written as follows:

\begin{equation} \label{eq:12}
X_{real} = \begin{bmatrix}
x^{real}_{11} & \dots & x^{real}_{1K} \\
\vdots & x^{real}_{nk} & \vdots \\
x^{real}_{N1} & \dots & x^{real}_{NK}
\end{bmatrix}
, X_{real} \in \mathbb{R}^{N \times K},
\end{equation}

\begin{equation} \label{eq:13}
X_{imag} = \begin{bmatrix}
x^{imag}_{11} & \dots & x^{imag}_{1K} \\
\vdots & x^{imag}_{nk} & \vdots \\
x^{imag}_{N1} & \dots & x^{imag}_{NK}
\end{bmatrix}
, X_{imag} \in \mathbb{R}^{N \times K},
\end{equation}

\begin{equation} \label{eq:14}
X = X_{real} + i \cdot X_{imag}, X \in \mathbb{C}^{N \times K}.
\end{equation}

In the basis-coefficients-solving module, the real and imaginary parts of the basis coefficients are embedded as two sets of weights between the input layer and first hidden layer, which are expressed in red and green, respectively, in Figure 2. For the first hidden layer, the activation function of each neuron is simply the linear function , which is omitted in the following equations. Note that there is a set of mirrors of $X_{real}$ and $X_{imag}$ for the complex operation. The architecture of the basis-coefficients-solving layer is determined by the complex algorithm as follows:

\begin{equation} \label{eq:15}
\hat{U} = \Psi \hat{X} = (\Psi_{real} \hat{X}_{real} - \Psi_{imag} \hat{X}_{imag}) + i \cdot (\Psi_{real} \hat{X}_{imag} + \Psi_{imag} \hat{X}_{real}), \hat{U} \in \mathbb{C}^{N \times K}.
\end{equation}

The masking module is designed to sample the reconstructed signal matrix by as follows:

\begin{equation} \label{eq:16}
\hat{Y} = P \odot \hat{U}, \hat{Y} \in \mathbb{C}^{N \times K},
\end{equation}

\noindent where $\hat{Y}$ denotes the masked reconstructed signals. In real-world applications, the masking matrices embedded in the transmitter and receiver are identical. Then, in the output layer, the target is the given signals $\hat{Y}$.

The loss function of the network is twofold: one is the mean-squared error (MSE) that measures the error between the masked reconstructed signals $\hat{Y}$ and the actual sampled signals $Y$; another is the $l_{\text{1}}$ regularizer that applys penalties on the weights $X_{real}$ and $X_{imag}$ during optimization. The loss function is given as

\begin{equation} \label{eq:17} \begin{split}
L &= \frac{1}{K} \sum_{n=1}^N \sum_{k=1}^K (y_{nk} - \hat{y}_{nk})^2 + 
\frac{\mu}{2} (\Vert X_{real} \Vert_1 + \Vert X_{imag} \Vert_1) \\
&= \frac{1}{K} \sum_{n=1}^N \sum_{k=1}^K (y_{nk} - \hat{y}_{nk})^2 + 
\frac{\mu}{2} \sum_{n=1}^N \sum_{k=1}^K (\vert x_{nk}^{real} \vert + \vert x_{nk}^{imag} \vert),
\end{split} \end{equation}

\noindent where the lower-case variables denote the elements of corresponding matrix. So far, the feed-forward procedure has been established. In this work, we use the Adam~\citep{RN1} algorithm [equation (\ref{eq:18})] to optimize the random initialized basis coefficients due to its superior convergence performance by combining stochastic gradient descent with momentum (SGDM)~\citep{RN64} and root-mean-square propagation (RMSProp)~\citep{RN65}:

\begin{subequations} \label{eq:18}
\begin{alignat}{1}
u_X^{(t+1)} &= \beta_1 u_X^{(t)} + (1-\beta_1) \nabla_X L^{(t)}, \\
v_X^{(t+1)} &= \beta_2 v_X^{(t)} + (1-\beta_2) (\nabla_X L^{(t)})^2, \\
\hat{u}_X &= \frac{u_X^{(t+1)}}{1 - (\beta_1)^{t+1}}, \\
\hat{v}_X &= \frac{v_X^{(t+1)}}{1 - (\beta_2)^{t+1}}, \\
X^{(t+1)} &= X^{(t)} - \eta \frac{\hat{u}_X}{\sqrt{\hat{v}_X} + \varepsilon},
\end{alignat}
\end{subequations}

\noindent where $X$ refers to both $X_{real}$ and $X_{imag}$ for simplicity, $t$ denotes the step of iteration, $u_X^{(t)}$ is the gradient with momentum of $X$, $v_X^{(t)}$ the gradient with second momentum of $X$, $\nabla_X L^{(t)}$ the gradient of $L^{(t)}$ with respect to $X$, and $\beta_1, \beta_2$ are the respective hyperparameters used to weight the momentum and second momentum terms.

Once we obtain the reconstructed signals, the reconstruction error between the original and reconstructed signals is calculated by

\begin{equation} \label{eq:19}
\xi = \vert \frac{\Vert U - U_{rec} \Vert_2}{\Vert U \Vert_2} \vert_{col},
\end{equation}

\noindent where $\vert \cdot \vert_{col}$ denotes column-wise operation, i.e., the norm is separately calculated for each channel; therefore, the reconstruction error $\xi$ is a row vector of dimension $1 \times K$.


\section{Examples}

\subsection{Examples using simple synthetic signals}

To validate the performance of the proposed CS-embedded network for data reconstruction, numerical simulations were conducted on multiple sinusoidal waves. First, five sinusoidal waves of frequency 10, 20, 30, 40, and 50 Hz were generated at a sampling frequency ; then, a superposed wave was synthesized from the first five waves by linear superposition. All six waves were sampled at by the masking module of the network; therefore, the equivalent sampling frequency is 80 Hz, which is already lower than twice the maximum predominant frequency (50 Hz) of signals 5 and 6, thus breaking the Nyquist-Shannon sampling theorem. The specific expressions of each simulated signal are given in equations (\ref{eq:19}). A Fourier-basis matrix of dimension is used as the input; correspondingly, the original signal matrix is split into four matrices of dimension (i.e., 5.12 s in duration) as the targets:

\begin{equation} \label{20}
\begin{split}
T &= 20.48 \text{ s}, f_s = 400 \text{ Hz}, \\
t &\in [0,T], dt = 1/f_s = 0.0025 \text{ s}, \\
s_n &= A_n \cos{(2 \pi \cdot f_n \cdot t + phase_n)}. \\
\end{split}
\end{equation}

\begin{equation*}
\begin{cases}
s_1 = \cos{(2\pi \cdot 10 \cdot t)} \\
s_2 = 0.1 \cdot \cos{(2\pi \cdot20 \cdot t + 0.1 \pi)} \\
s_3 = 0.3 \cdot \cos{(2\pi \cdot30 \cdot t + 0.3 \pi)} \\
s_4 = 0.5 \cdot \cos{(2\pi \cdot40 \cdot t + 0.5 \pi)} \\
s_5 = 0.7 \cdot \cos{(2\pi \cdot50 \cdot t + 0.7 \pi)} \\
s_6 = s_1 + s_2 + s_3 + s_4 + s_5
\end{cases}
\end{equation*}

\begin{figure}[H]
    \begin{center}$
    \begin{array}{cc}
    \includegraphics[width=.45\textwidth]{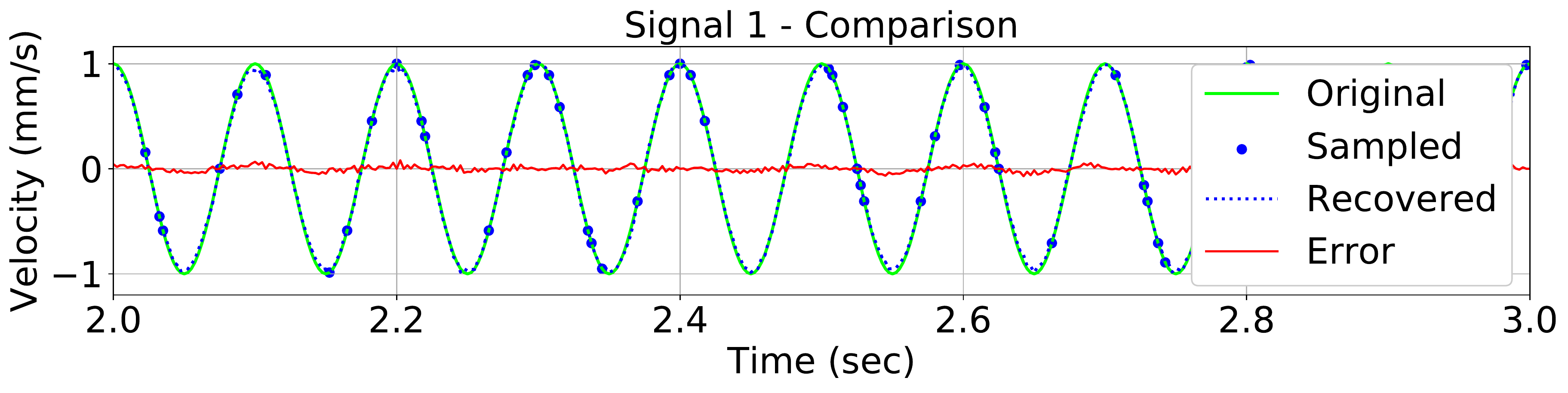} &
    \includegraphics[width=.45\textwidth]{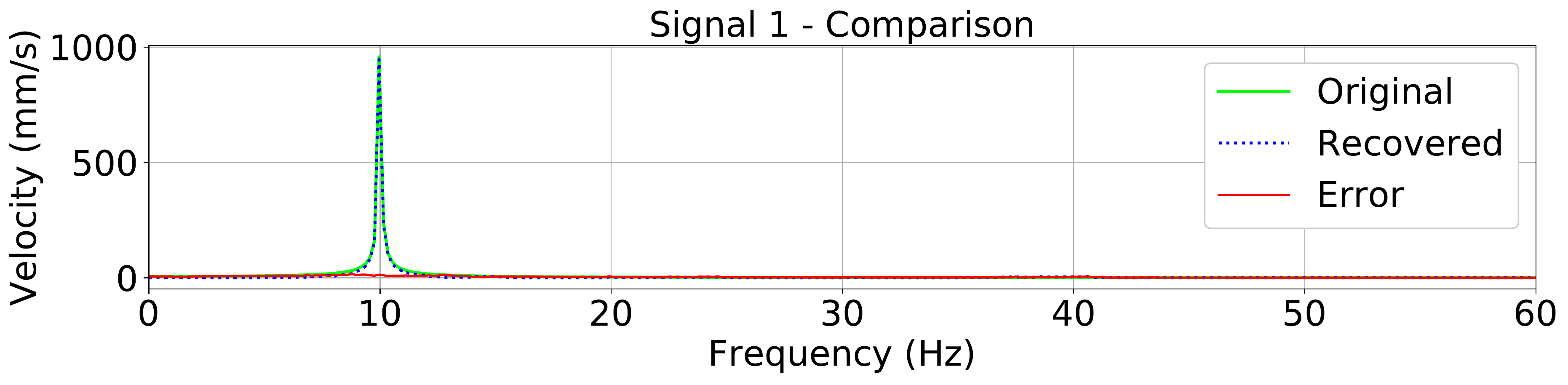} \\
    \includegraphics[width=.45\textwidth]{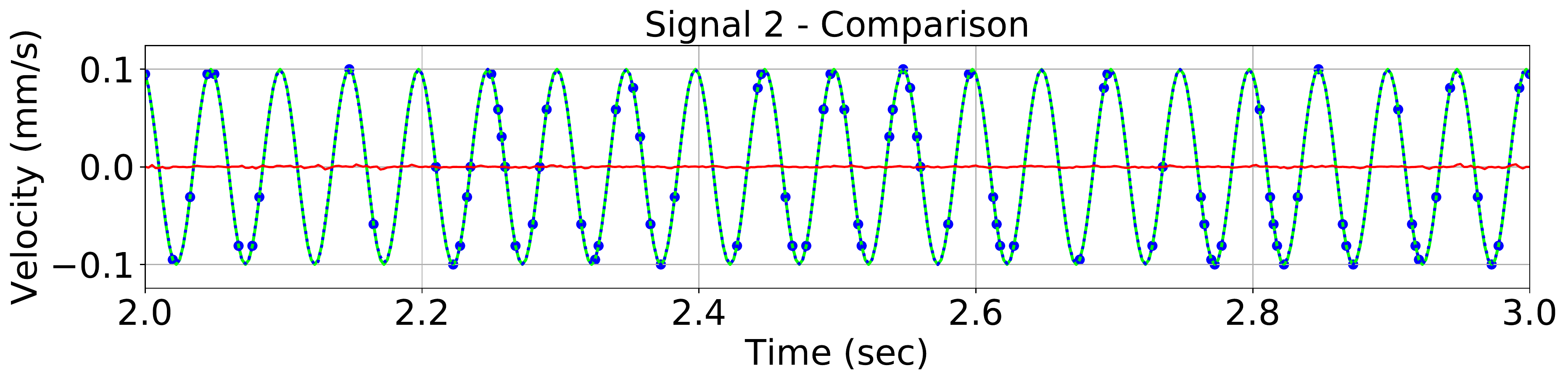} &
    \includegraphics[width=.45\textwidth]{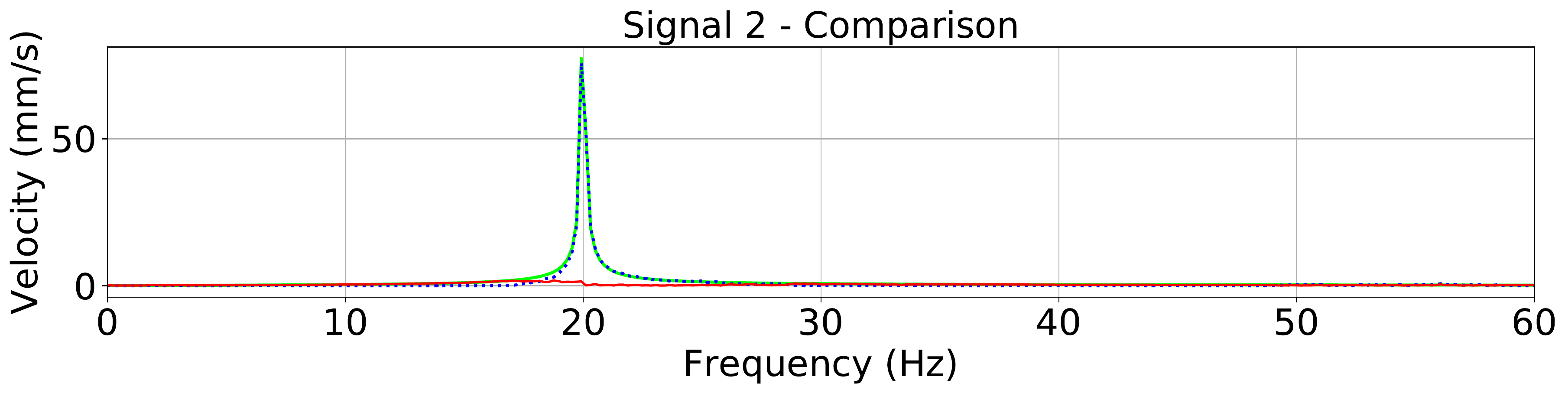} \\
    \includegraphics[width=.45\textwidth]{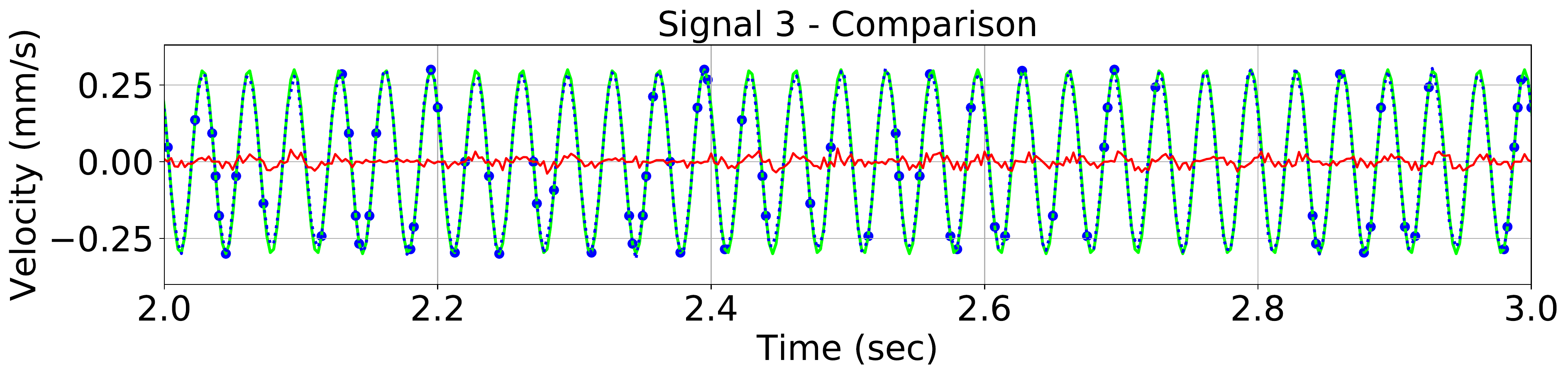} &
    \includegraphics[width=.45\textwidth]{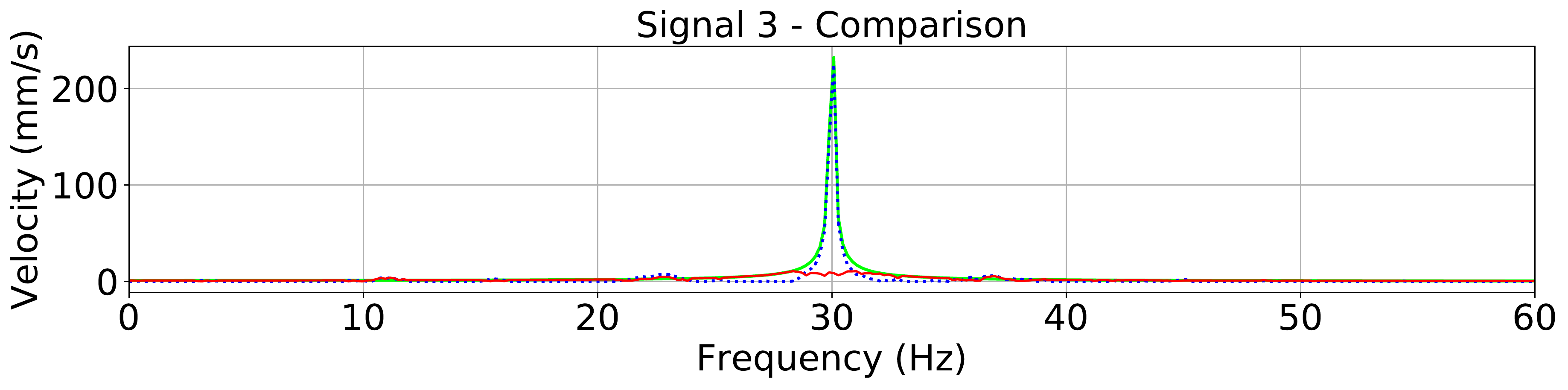} \\
    \end{array}$
    \end{center}
    \label{fig:fig_3_no_caption}
\end{figure}

\begin{figure}[H]
    \begin{center}$
    \begin{array}{cc}
    \includegraphics[width=.45\textwidth]{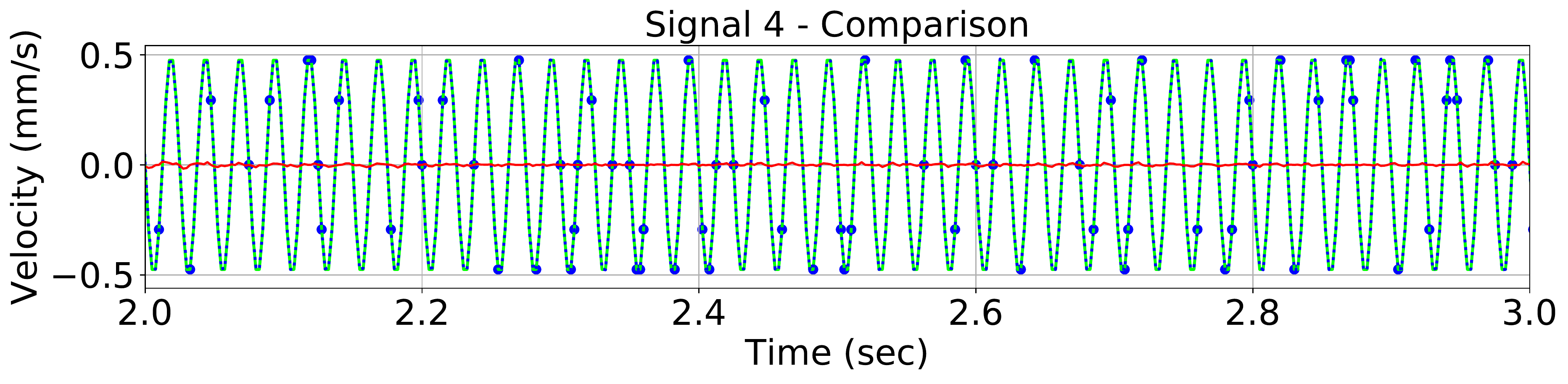} &
    \includegraphics[width=.45\textwidth]{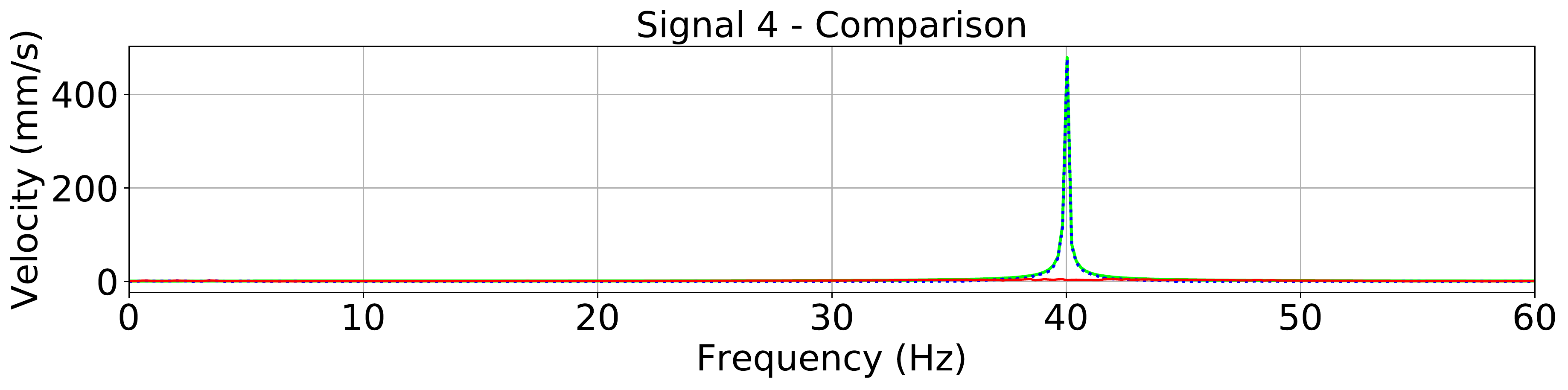} \\
    \includegraphics[width=.45\textwidth]{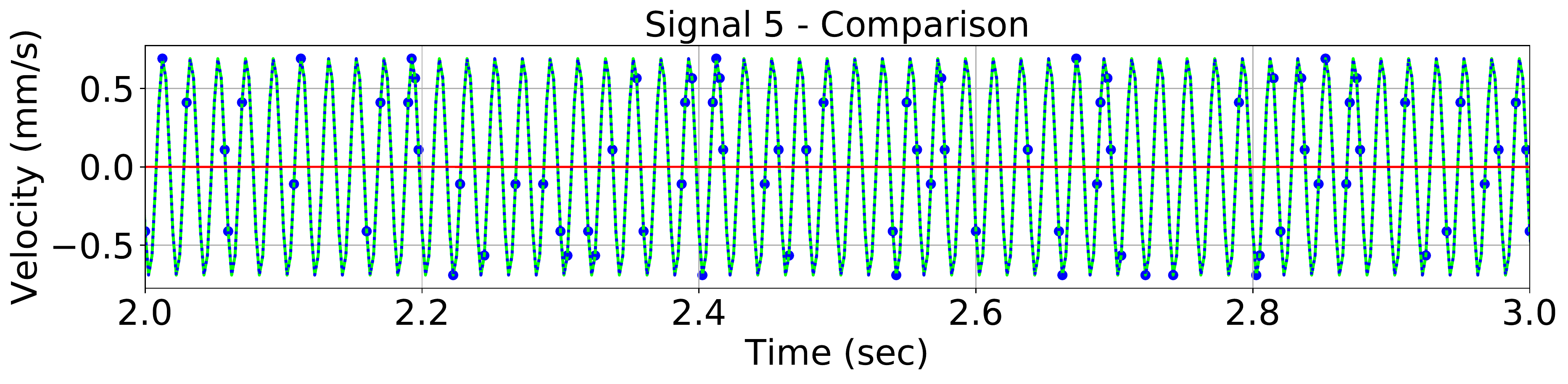} &
    \includegraphics[width=.45\textwidth]{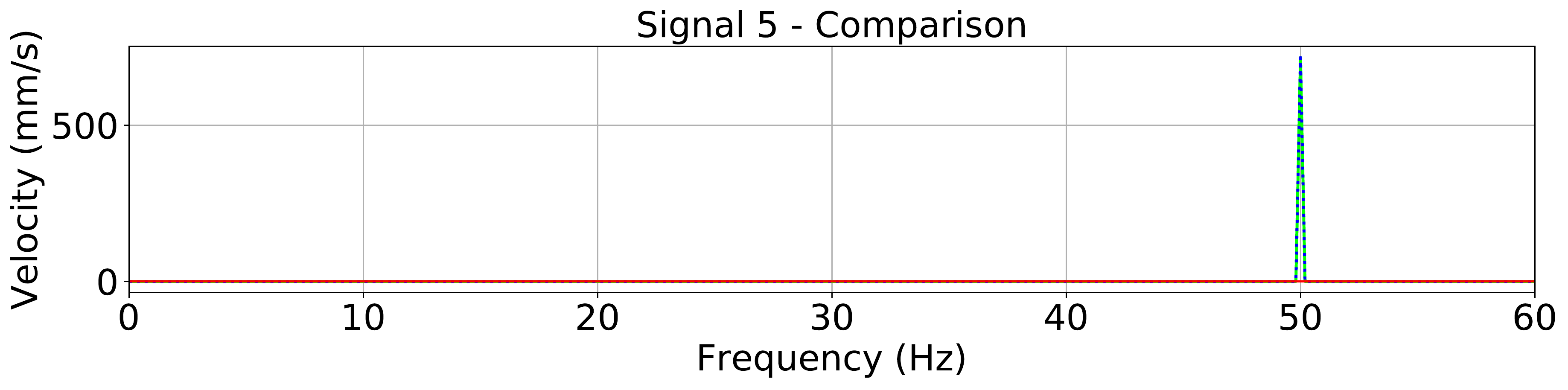} \\
    \includegraphics[width=.45\textwidth]{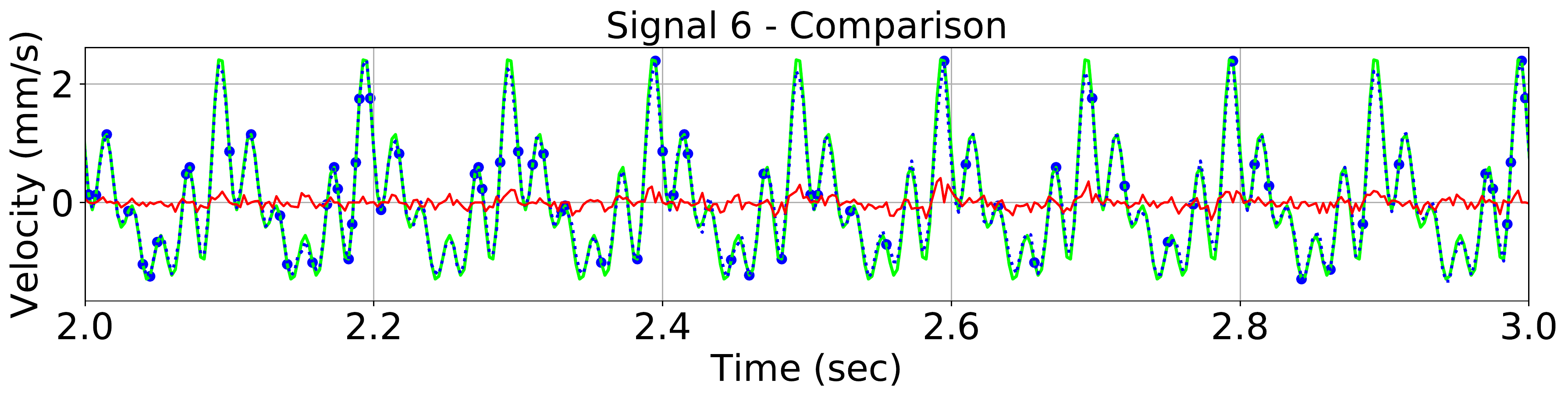} &
    \includegraphics[width=.45\textwidth]{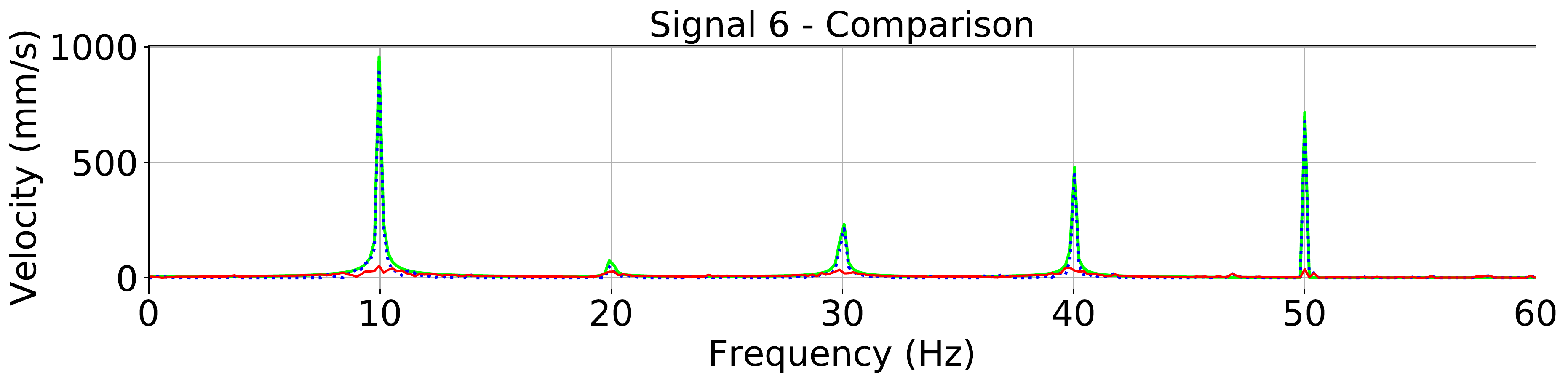} \\
    \end{array}$
    \end{center}
    \caption{Comparison between original signals and reconstructed signals (time responses are displayed for 1 s for visibility): left column: time domain; right column: frequency domain}
    \label{fig:fig_3}
\end{figure}

\begin{figure}[H]
\centering
\begin{subfigure}{0.48\textwidth}
\centering
\includegraphics[clip, trim=0.3cm 0.3cm 0.3cm 0.3cm, width=0.9\textwidth]{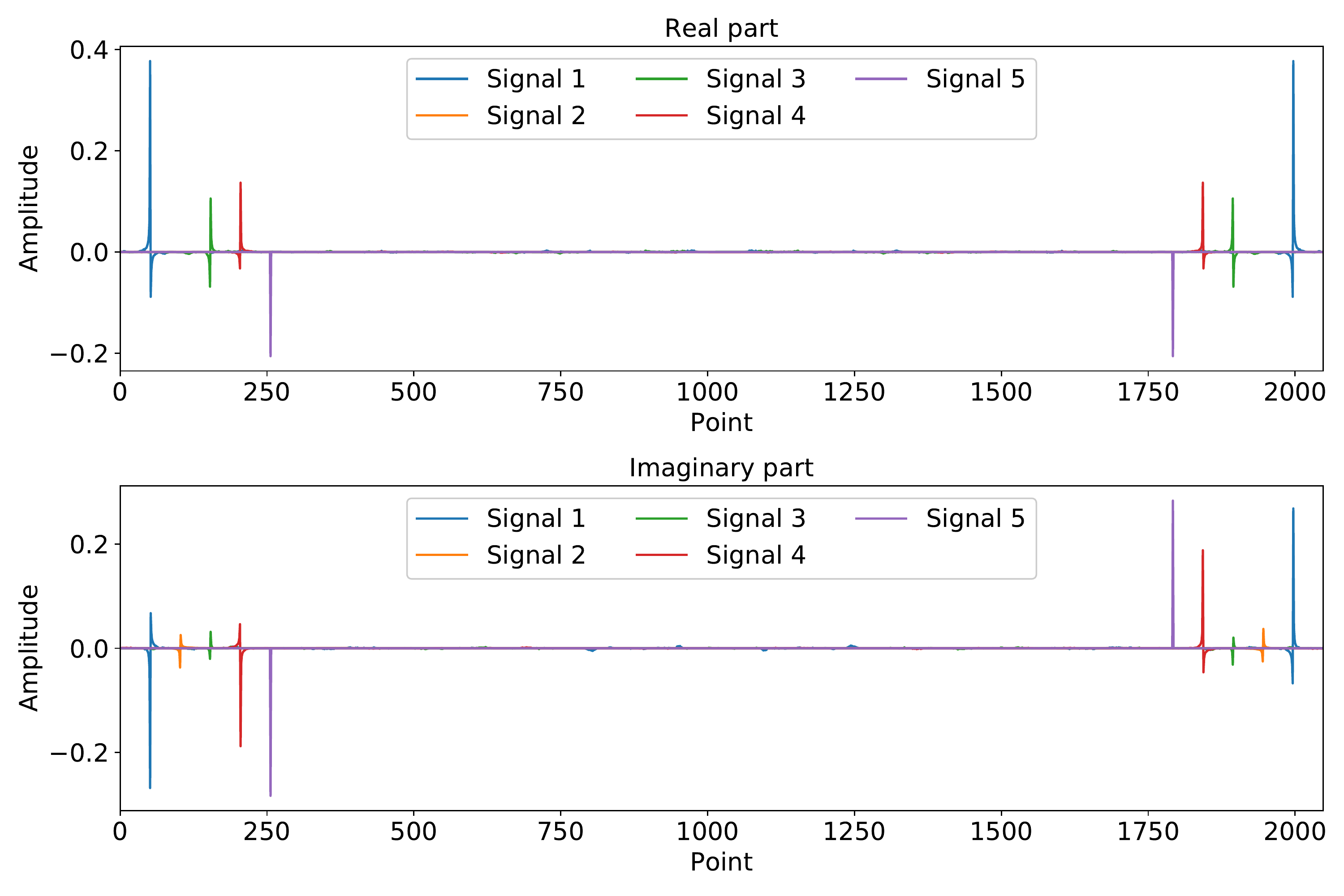}
\caption{}
\label{fig:4-a}
\end{subfigure}
\begin{subfigure}{0.48\textwidth}
\centering
\includegraphics[clip, trim=0.3cm 0.3cm 0.3cm 0.3cm, width=0.9\textwidth]{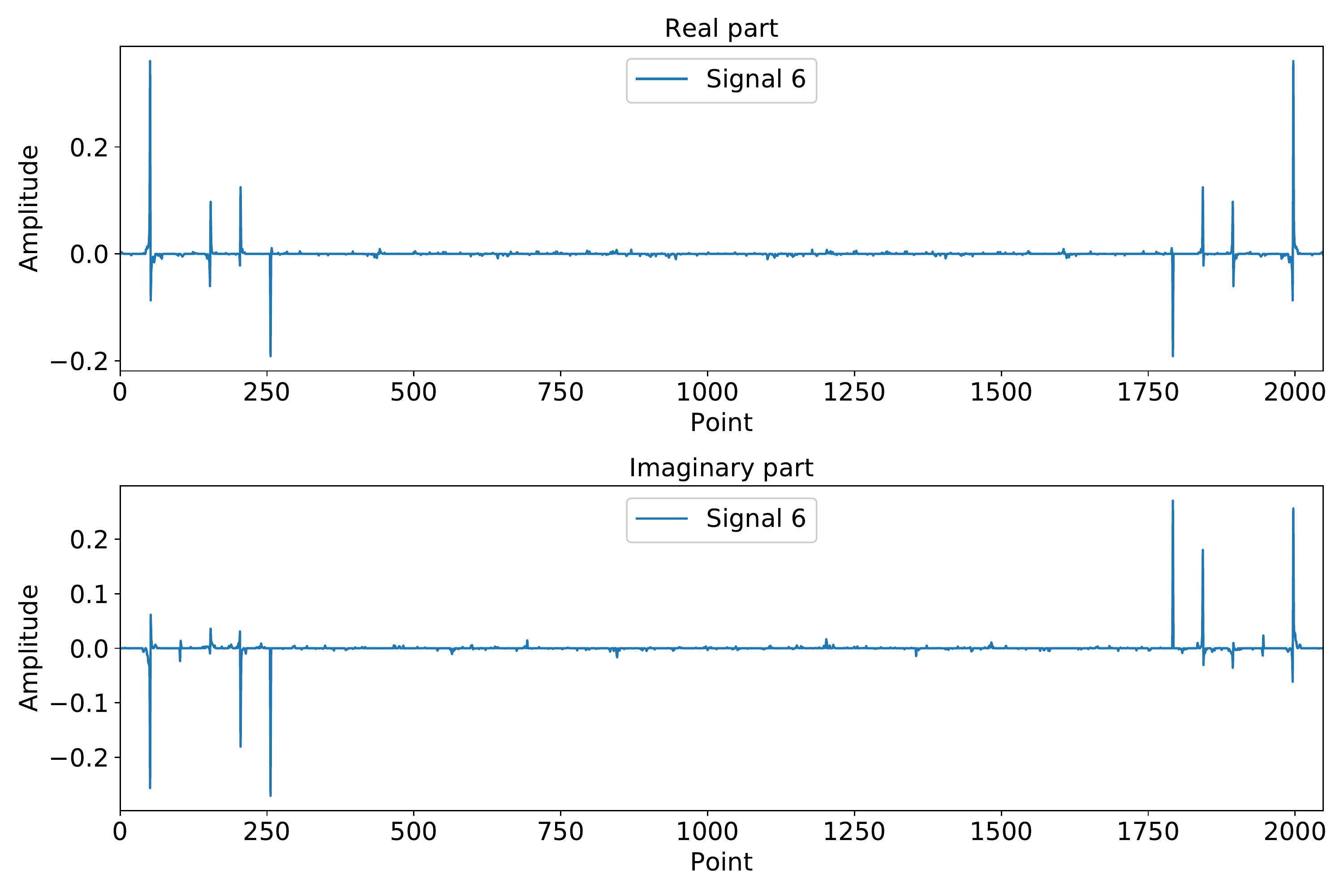}
\caption{}
\label{fig:4-b}
\end{subfigure}
\caption{Reconstructed basis coefficients of each signal: (a) signals 1-5 ; (b) signal 6}
\label{fig:fig_4}
\end{figure}

The comparison results in Figure 3 show that although the sampled points are quite sparse, all six signals are well reconstructed both in the time and frequency domains. Signals 5 and 6 even achieved better reconstruction compared with the other signal channels. Moreover, without any constraints except for l1-norm minimization for $X$, in Figure 4 the real and imaginary parts of the reconstructed basis coefficients show the fine characteristics of even and odd symmetry, respectively, which agree on the nature of the Fourier basis.

\subsection{Examples using field test wireless data of a long-span bridge}

Next, we used more complex data collected from a field test on Xiamen Haicang Bridge to evaluate the performance of the proposed approach. As shown in Figure 5(a), the bridge is a steel-box-girder suspension bridge with a span distribution of 230 m+648 m+230 m, two towers with a height of 140 m, and a bridge deck 32 m wide. To avoid disturbing the normal traffic, the test was carried out at midnight. The test schemes are shown in Figure 6. Specifically, the tests were repeated nine times and tested a total of 62 test points. Test 1 had seven test points (Nos. 1–6 and No. 26); wireless sensors were placed on the seven test points to measure the vibration data. The time duration of each test was 20 min. After Test 1 finished, the wireless sensors were moved for Test 2, which also included seven test points (Nos. 7–12 and No. 26). Considering the effective wireless-data-transmission distances, test point No. 26 was selected as the reference point for all nine tests. Tests 3–9 were conducted in the same way. The field test used nine commercial wireless velocity sensors as shown in Figure 5(b); the sampling frequency for data acquisition was 100 Hz.

\begin{figure}[t]
\centering
\begin{subfigure}{0.45\textwidth}
\centering
\includegraphics[height=4cm]{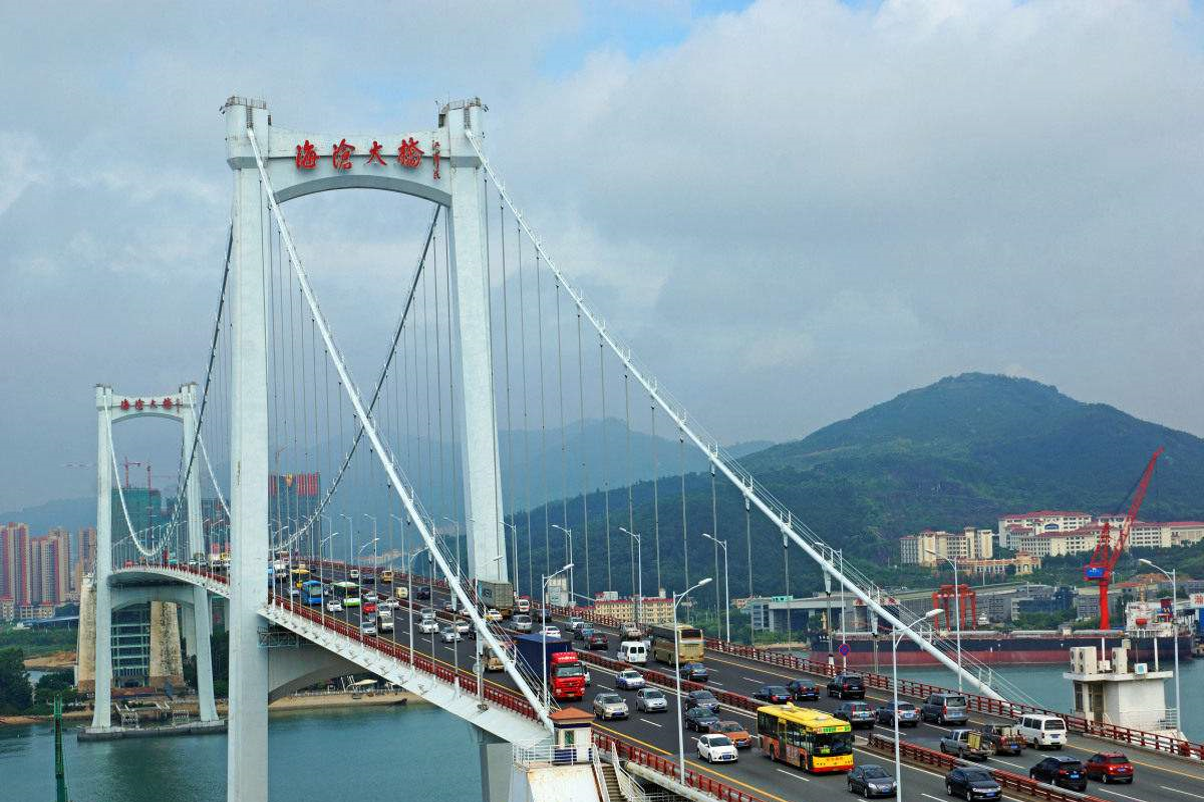} 
\caption{}
\label{fig:5-a}
\end{subfigure}
\hfill
\begin{subfigure}{0.45\textwidth}
\centering
\includegraphics[height=4cm]{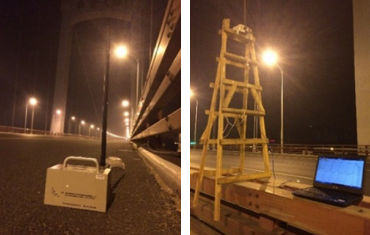}
\caption{}
\label{fig:5-b}
\end{subfigure}
\caption{(a) Xiamen Haicang Bridge; (b) wireless sensor and base station}
\label{fig:fig_5}
\end{figure}

\begin{figure}[htb]
\centering
\includegraphics[width=0.75\textwidth]{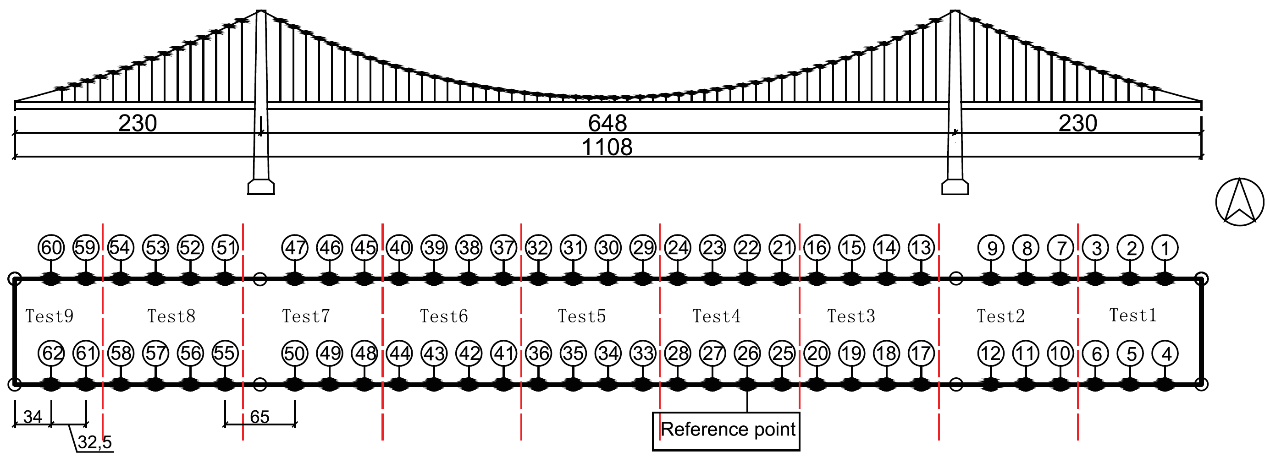}
\label{fig:fig_6}
\caption{Placement of test points}
\end{figure}

We simulate the CS-sampling procedure at nine different sampling ratios ranging from 10\% to 50\% in steps of 5\% for all 62 points. The length of the signal slice and the dimension of the basis matrix are both set to 2048. Table~\ref{tab:table_1} lists the schedule of parameters during optimization. It is found that the real part of the signal is slower to converge in the optimization procedures, so the loss weight pair is introduced in the output layer to magnify the loss individually, which mainly speeds up the training of the real part. Figure 7 and Figure 8 demonstrate the typical signal-reconstruction process (Test 2, point 26, $\gamma=0.3$). Specifically, Figure 7 visualizes the reconstructed time response at epochs 100, 150, 200, 400, and 600. It is shown that the reconstructed response remains almost zero after 100 epochs of training; then, it gradually converges to the original. Figure 8 shows the loss history of $X_{real}$ and $X_{imag}$, as well as the evolution of the shapes of $X_{real}$ and $X_{imag}$.

\renewcommand\arraystretch{1.2}
\begin{table}[hbt]
 \caption{Schedule of parameters during optimization}
  \begin{center}
  \begin{tabular}{ccccc}
    \hline
    \multirow{2}{*}{Epoch} & \multirow{2}{*}{\shortstack{Learning \\ rate}} & \multicolumn{2}{c}{Loss weight} & \multirow{2}{*}{Batch size} \\
    \cline{3-4} 
    & & Real part & Imaginary part & \\
    \hline
    1-100   & \multirow{3}{*}{$1\mathrm{e}{-4}$}   & 1    & 1    & \multirow{6}{*}{128} \\
    101-200 &                                      & 128  & 1   \\
    201-300 &                                      & 256  & 1   \\ \cline{2-2}
    301-400 & \multirow{3}{*}{$1\mathrm{e}{-5}$}   & 1024 & 1   \\
    401-500 &                                      & 4096 & 512 \\
    501-600 &                                      & 8192 & 512 \\
    \hline
  \end{tabular}
  \end{center}
  \label{tab:table_1}
\end{table}

\begin{figure}[H]
\centering
\begin{subfigure}{0.85\textwidth}
\centering
\includegraphics[clip, trim=0cm 0.3cm 0cm 0cm, width=0.9\textwidth]{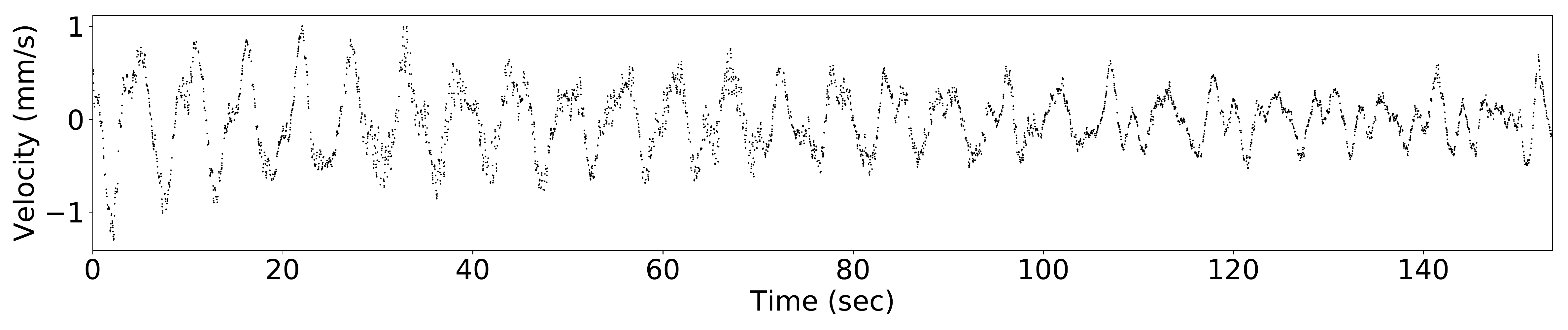}
\caption{}
\label{fig:7-a}
\end{subfigure}
\end{figure}

\begin{figure}[H]
\setcounter{figure}{6}
\centering
\begin{subfigure}{0.85\textwidth}
\setcounter{subfigure}{1}
\centering
\includegraphics[clip, trim=0cm 0.3cm 0cm 0cm, width=0.9\textwidth]{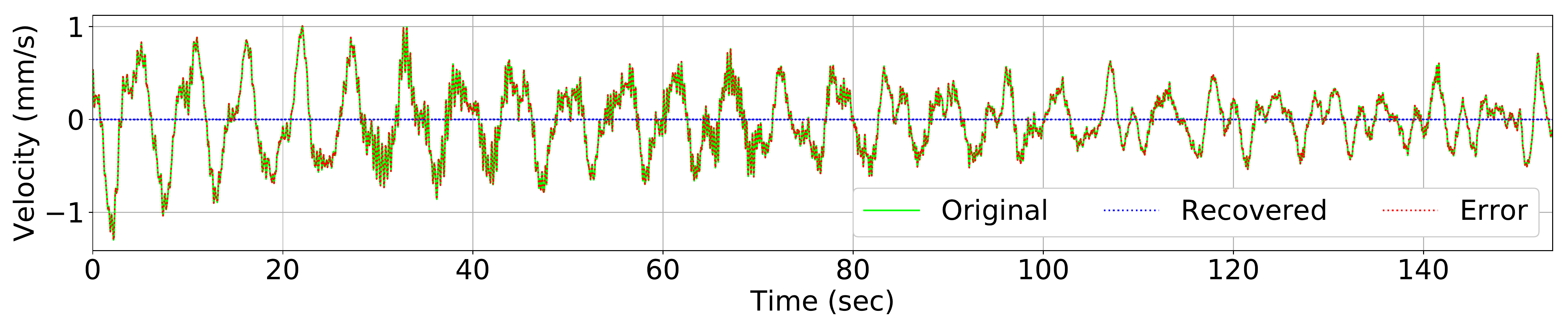}
\caption{}
\label{fig:7-b}
\end{subfigure}
\begin{subfigure}{0.85\textwidth}
\centering
\includegraphics[clip, trim=0cm 0.3cm 0cm 0cm, width=0.9\textwidth]{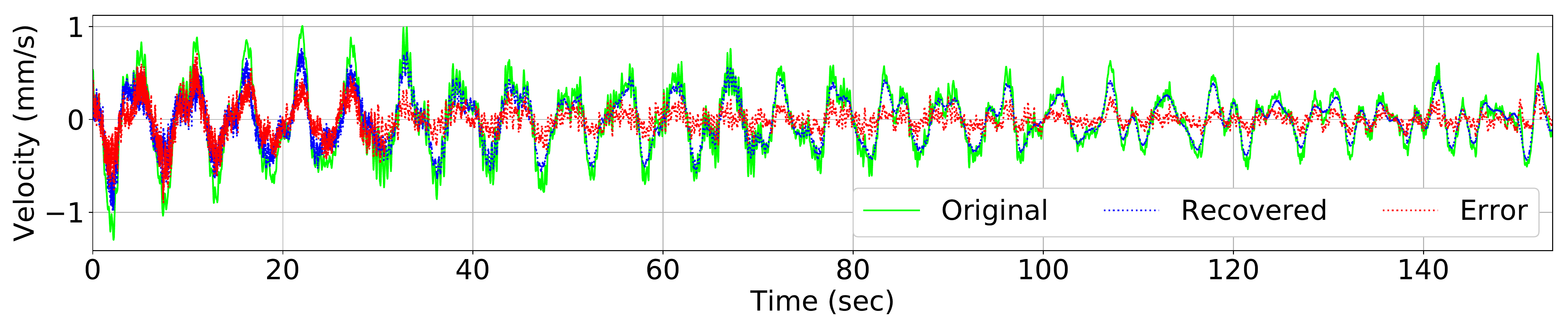} 
\caption{}
\label{fig:7-c}
\end{subfigure}
\begin{subfigure}{0.85\textwidth}
\centering
\includegraphics[clip, trim=0cm 0.3cm 0cm 0cm, width=0.9\textwidth]{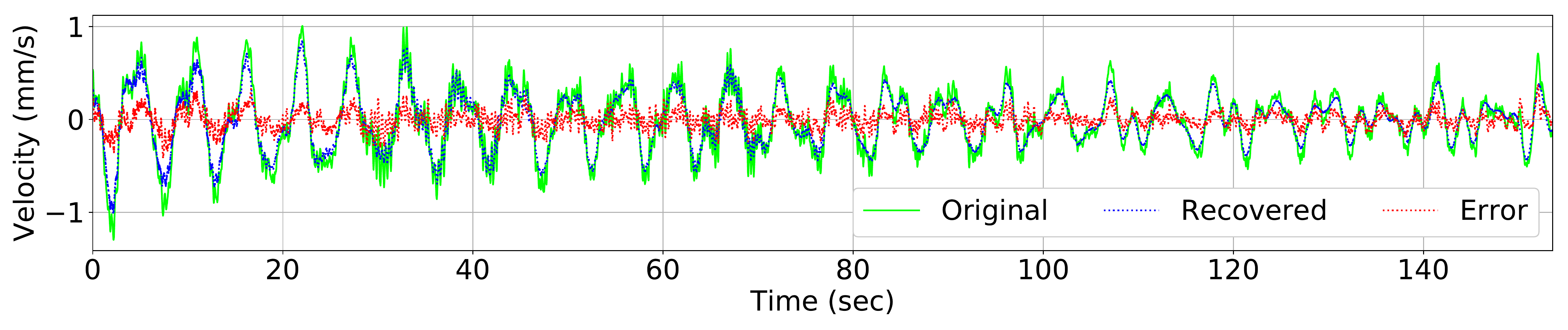} 
\caption{}
\label{fig:7-d}
\end{subfigure}
\begin{subfigure}{0.85\textwidth}
\centering
\includegraphics[clip, trim=0cm 0.3cm 0cm 0cm, width=0.9\textwidth]{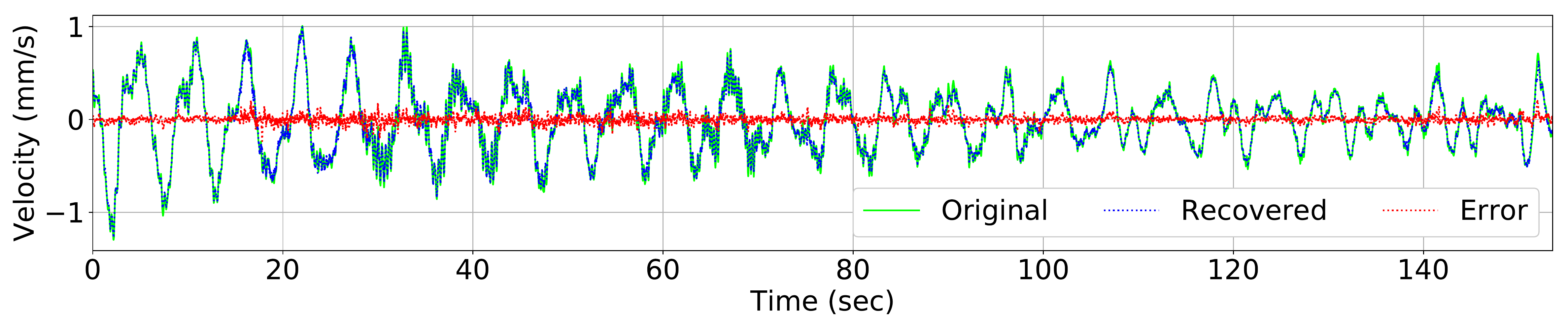} 
\caption{}
\label{fig:7-e}
\end{subfigure}
\begin{subfigure}{0.85\textwidth}
\centering
\includegraphics[clip, trim=0cm 0.3cm 0cm 0cm, width=0.9\textwidth]{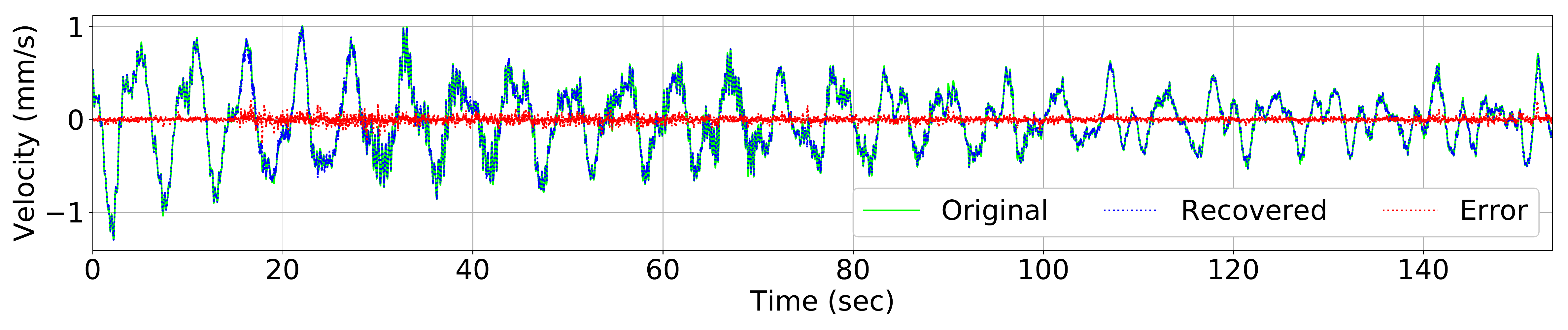} 
\caption{}
\label{fig:7-f}
\end{subfigure}
\caption{(a) Sampled data with sampling ratio 0.3 of Test 2, point 26; (b)–(f) typical signal-reconstruction process at epochs 100, 150, 200, 400, and 600, respectively.}
\label{fig:fig_7}
\end{figure}

\begin{figure}[H]
\centering
\begin{subfigure}[b]{.42\linewidth}
\includegraphics[clip, trim=0cm 1cm 0.5cm 0cm, width=\linewidth]{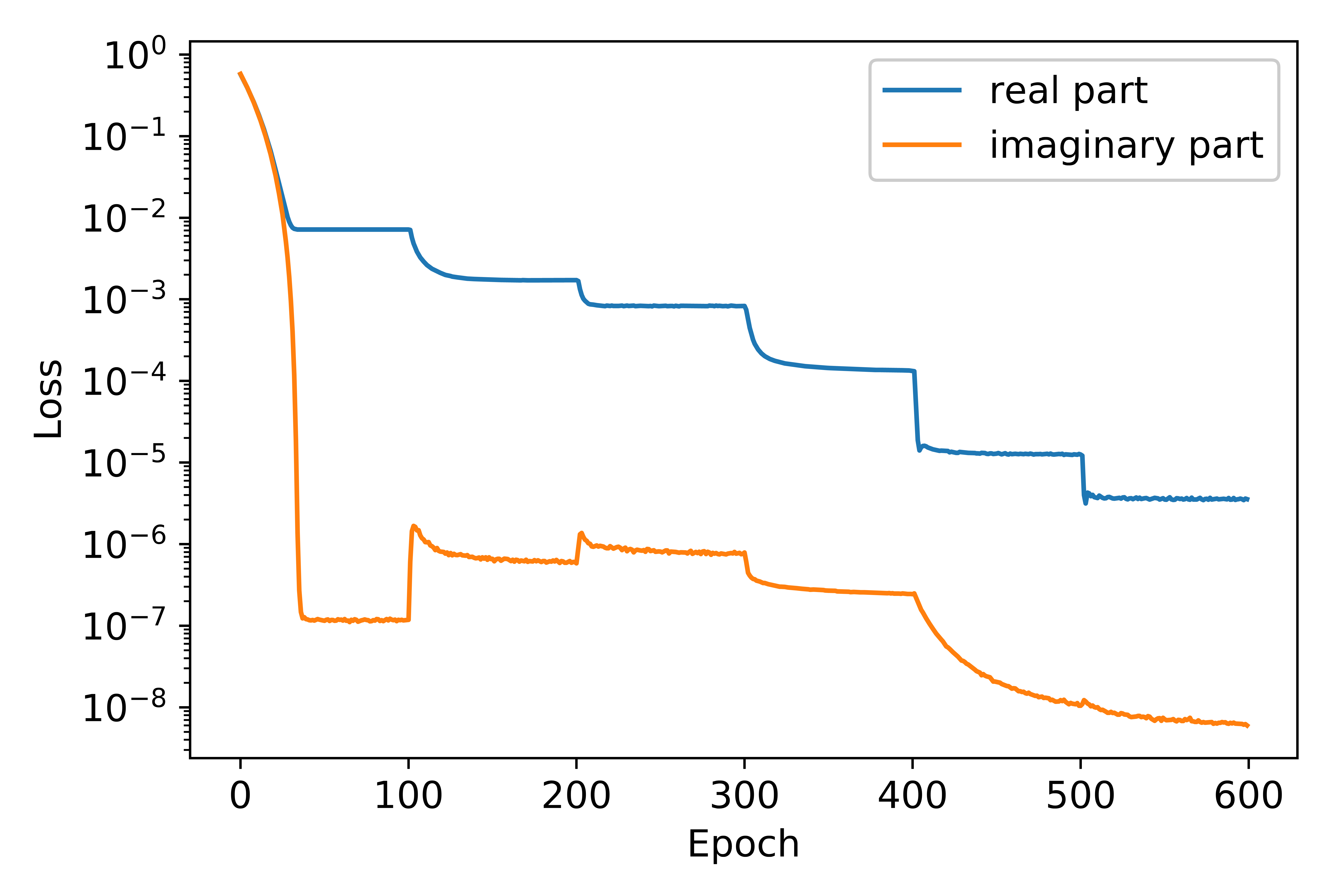}
\caption{}\label{fig:8-a}
\end{subfigure}
\end{figure}

\begin{figure}[H]
\setcounter{figure}{7}
\centering

\begin{subfigure}[b]{.45\linewidth}
\setcounter{subfigure}{1}
\includegraphics[clip, trim=1.8cm 1.5cm 0.5cm 0.5cm, width=\linewidth]{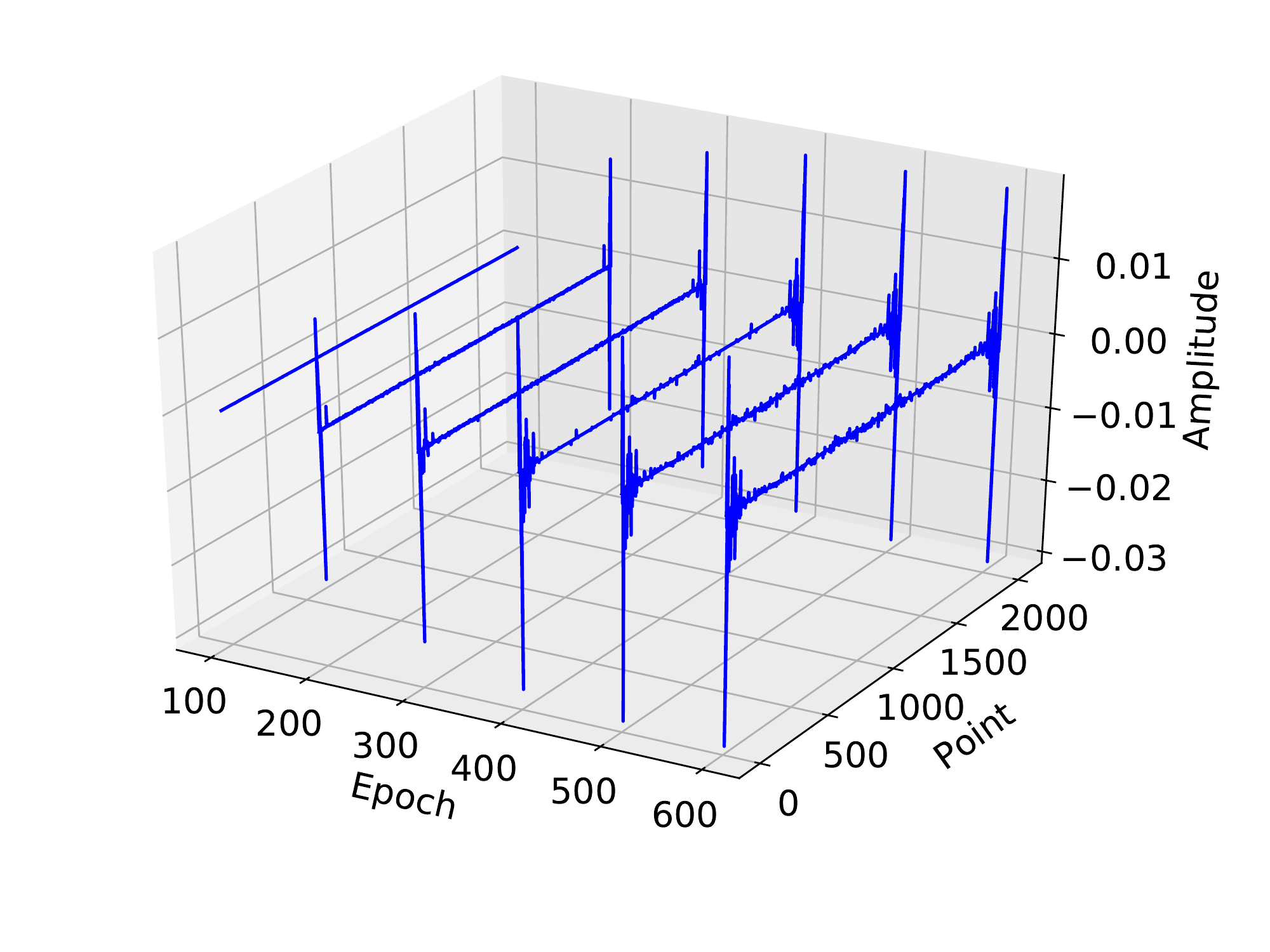}
\caption{}\label{fig:8-b}
\end{subfigure}
\begin{subfigure}[b]{.45\linewidth}
\includegraphics[clip, trim=1.8cm 1.5cm 0.5cm 0.5cm, width=\linewidth]{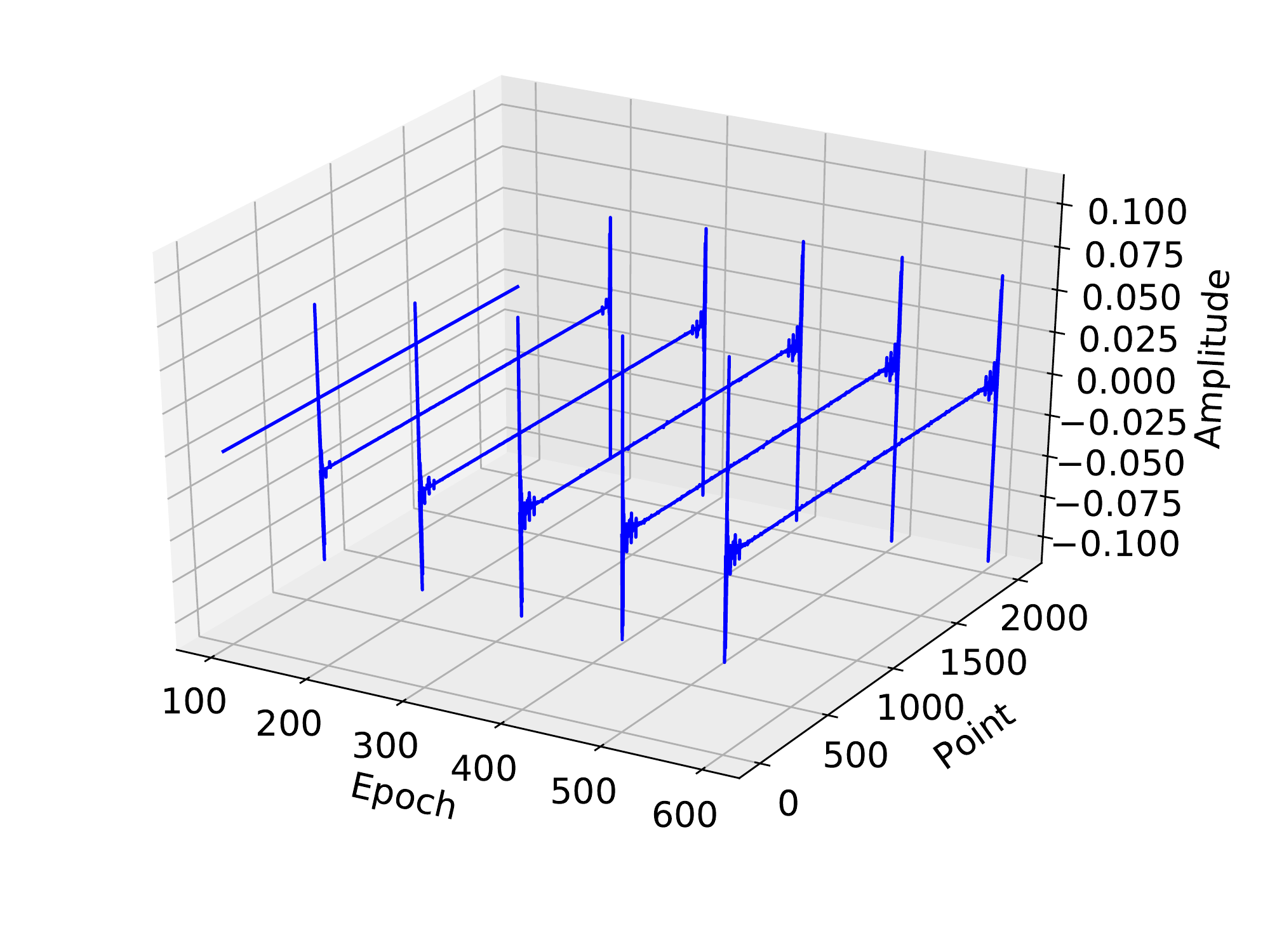}
\caption{}\label{fig:8-c}
\end{subfigure}
\caption{Optimization process of basis coefficients at sampling ratio 0.3 of Test 2, point 26: (a) loss history; (b) real part; (c) imaginary part}
\label{fig:fig_8}
\end{figure}

Nine grouped test results in Figure 9 show the consistent decrease of reconstruction error as the sampling ratio increases from 10\% to 50\% in steps of 5\%. At a sampling ratio of 0.3, the best reconstruction appears at Test 6, channel 7, which achieves an error of only 0.0474; meanwhile, the reconstruction errors of other signals are approximately 0.1; at a sampling ratio of 0.5, again, channel 7 of Test 6 accomplishes the minimal reconstruction error of 0.0196; other signals are reconstructed with errors ranging approximately between 0.05 and 0.1.

\begin{figure}[H]
\centering

\begin{subfigure}[b]{.38\linewidth}
\includegraphics[clip, trim=0cm 0.3cm 0cm 0.3cm, width=\linewidth]{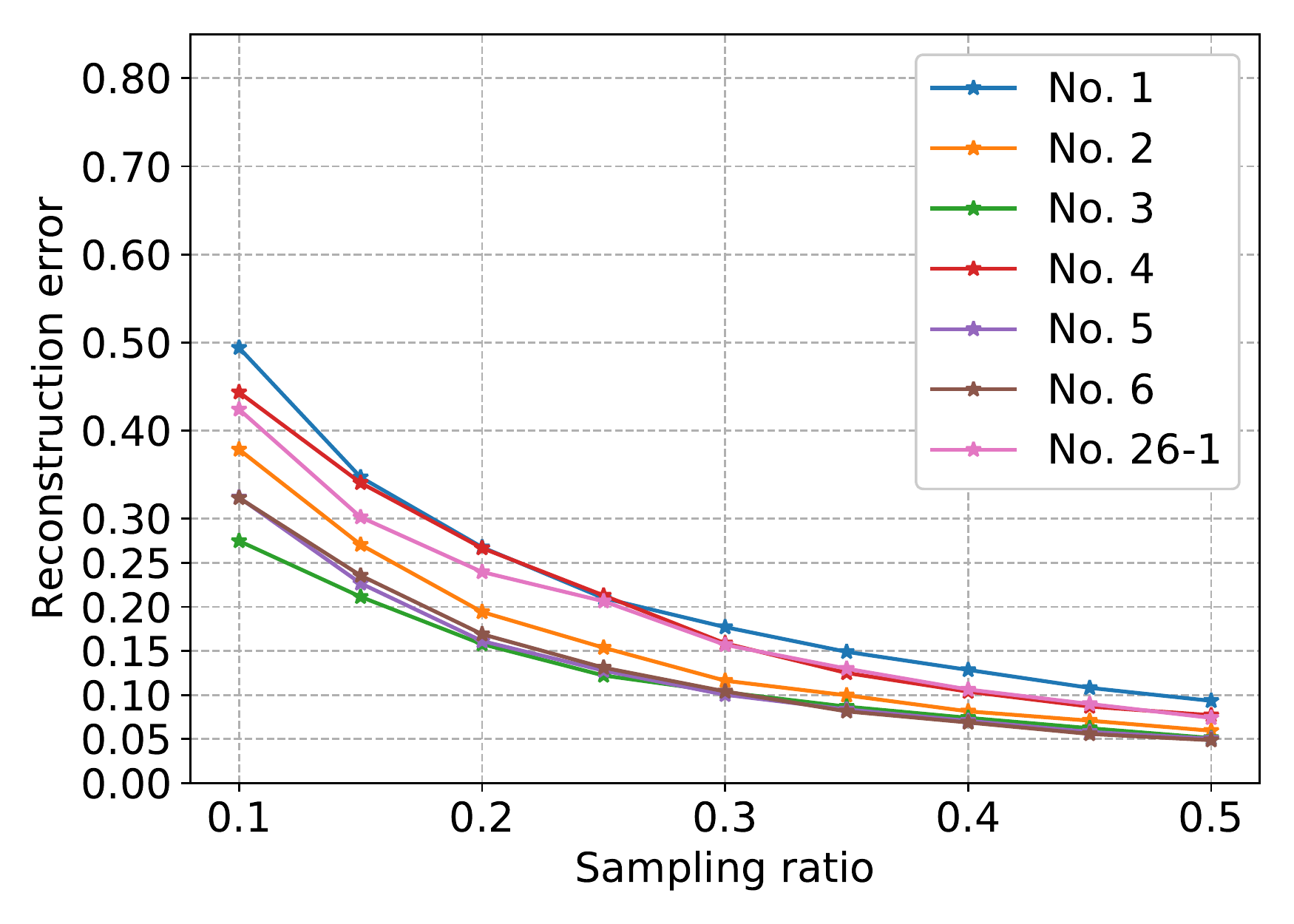}
\label{fig:9-1-1}
\end{subfigure}
\begin{subfigure}[b]{.38\linewidth}
\includegraphics[clip, trim=0cm 0.3cm 0cm 0.3cm, width=\linewidth]{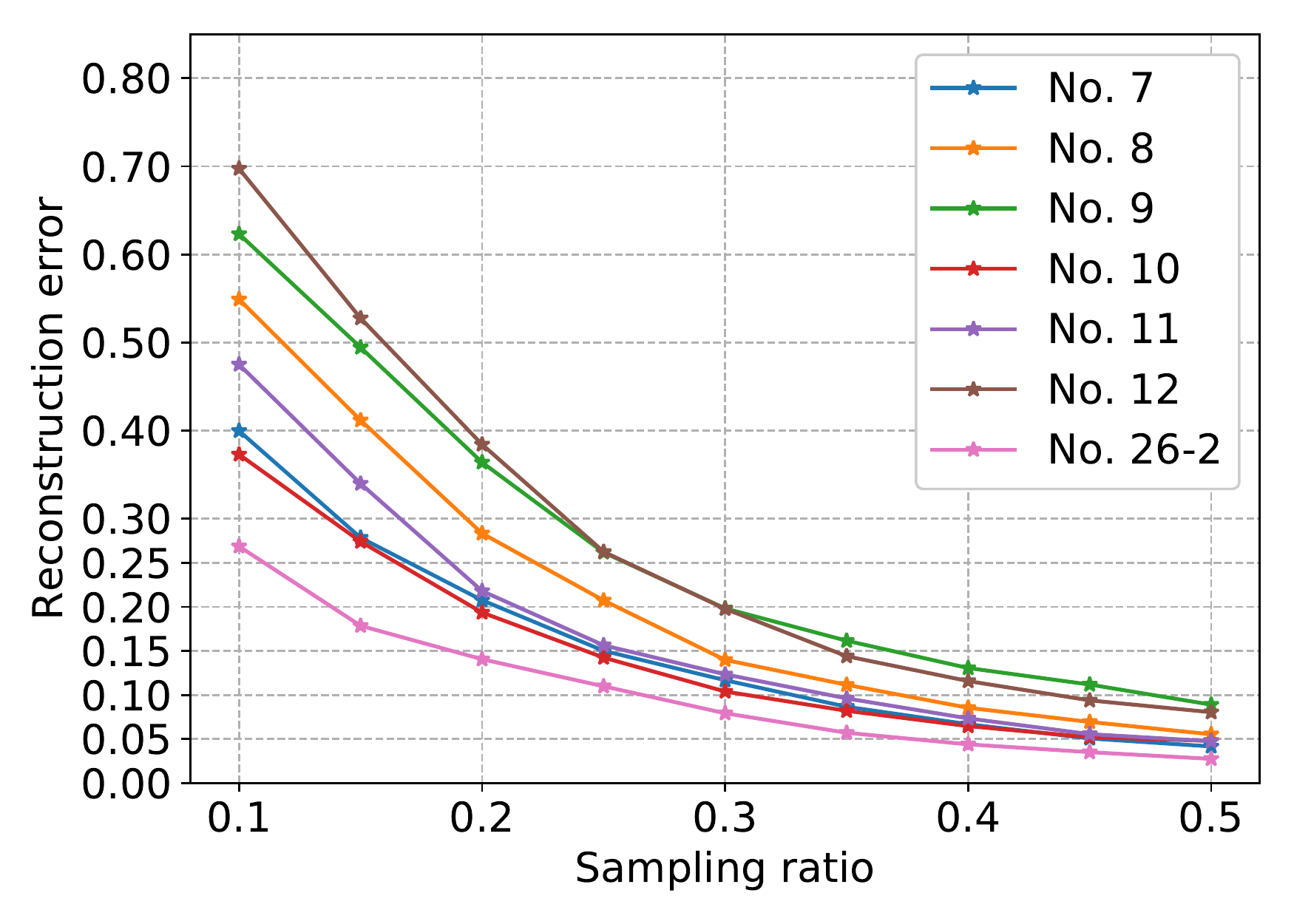}
\label{fig:9-1-2}
\end{subfigure}
\vspace{-1\baselineskip}

\begin{subfigure}[b]{.38\linewidth}
\includegraphics[clip, trim=0cm 0.3cm 0cm 0.3cm, width=\linewidth]{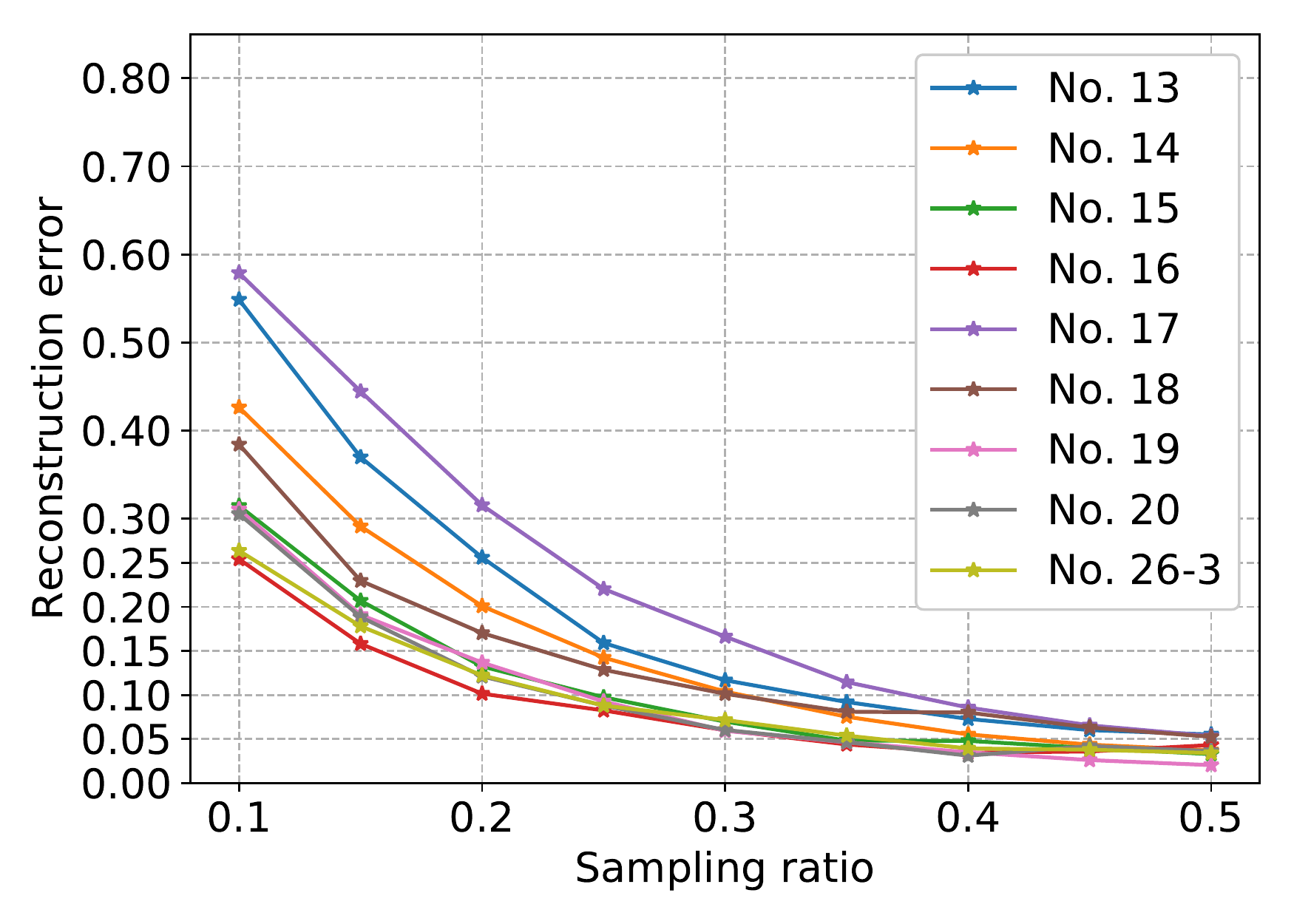}
\label{fig:9-2-1}
\end{subfigure}
\begin{subfigure}[b]{.38\linewidth}
\includegraphics[clip, trim=0cm 0.3cm 0cm 0.3cm, width=\linewidth]{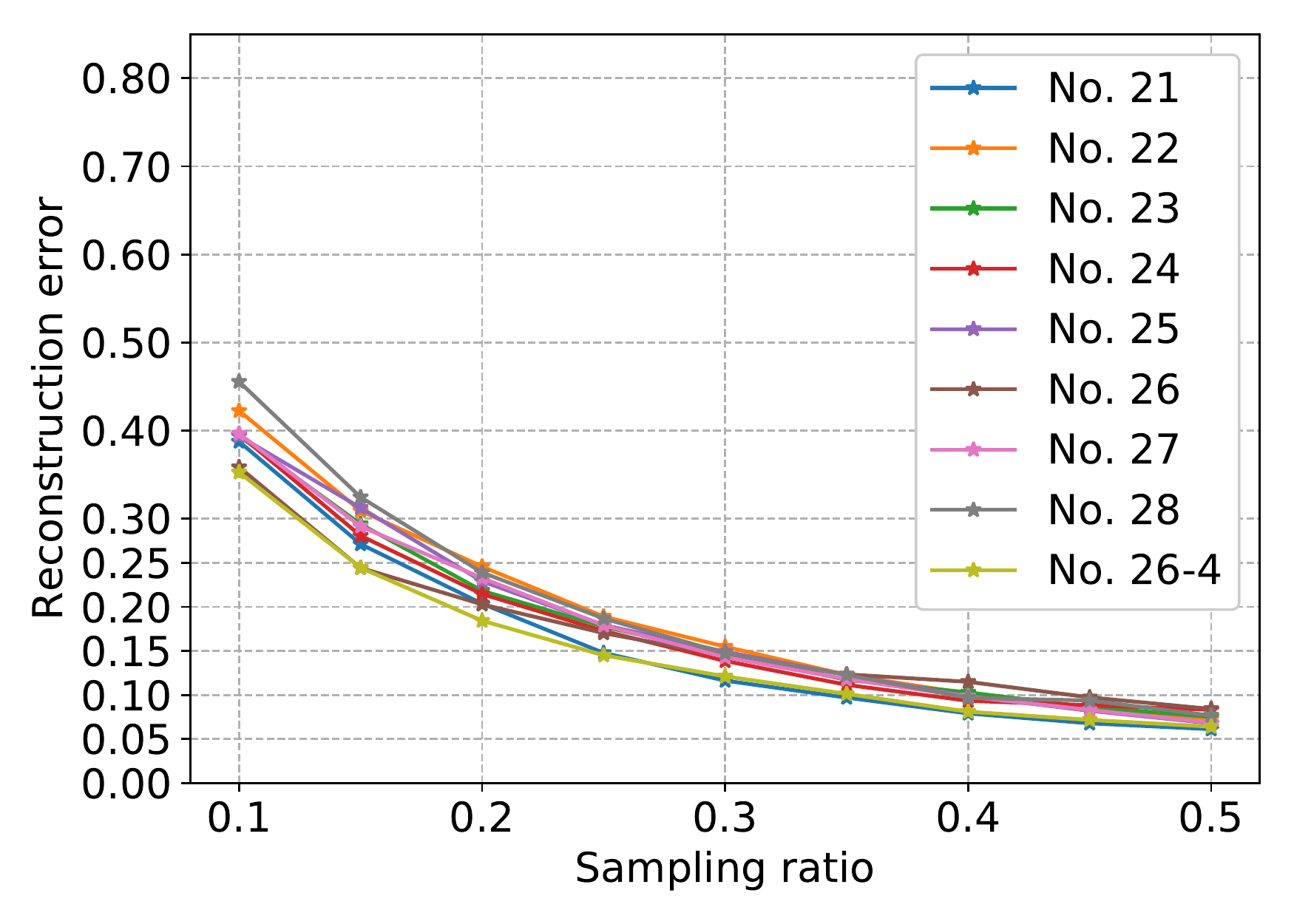}
\label{fig:9-2-2}
\end{subfigure}
\vspace{-1\baselineskip}

\begin{subfigure}[b]{.38\linewidth}
\includegraphics[clip, trim=0cm 0.3cm 0cm 0.3cm, width=\linewidth]{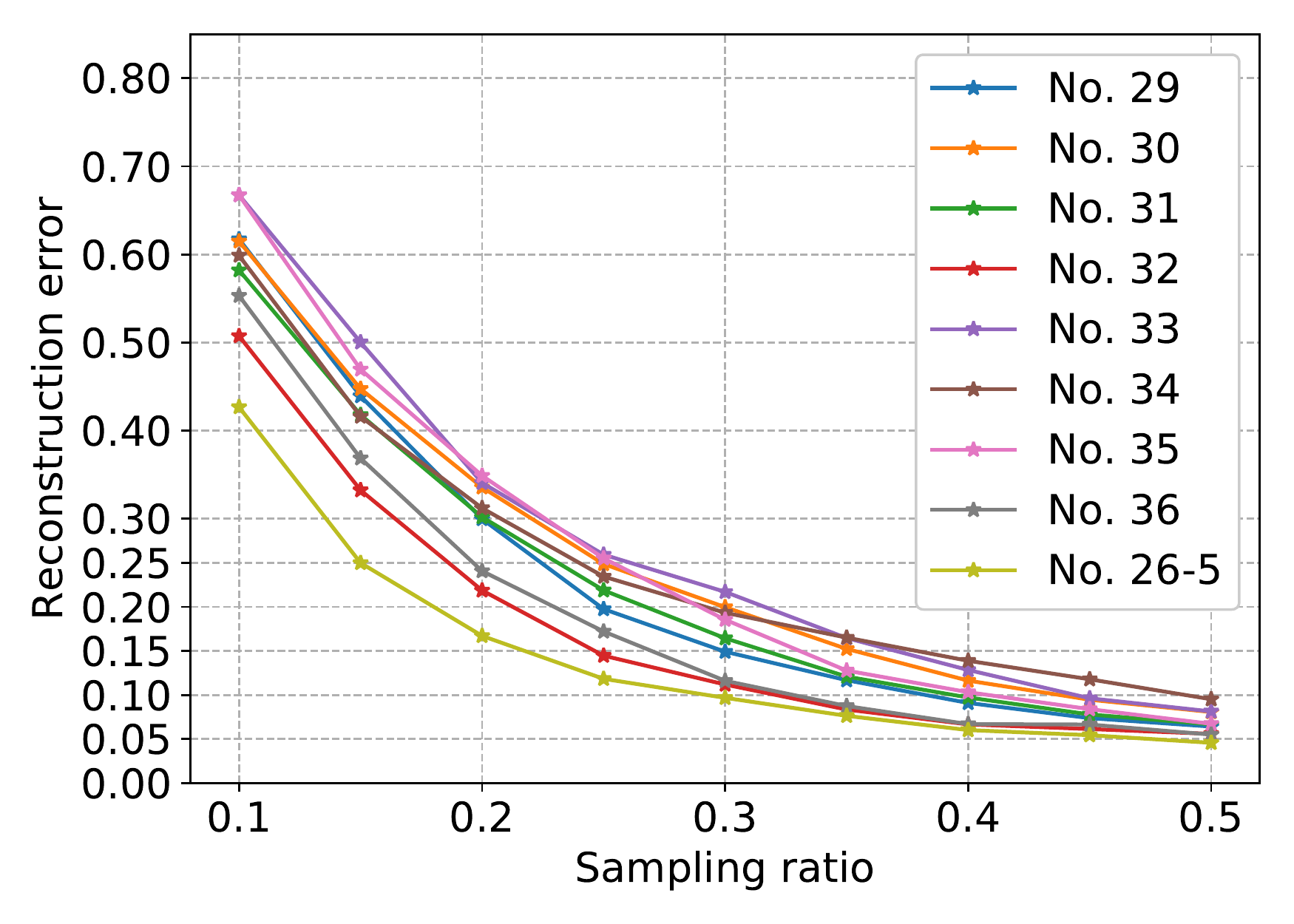}
\label{fig:9-3-1}
\end{subfigure}
\begin{subfigure}[b]{.38\linewidth}
\includegraphics[clip, trim=0cm 0.3cm 0cm 0.3cm, width=\linewidth]{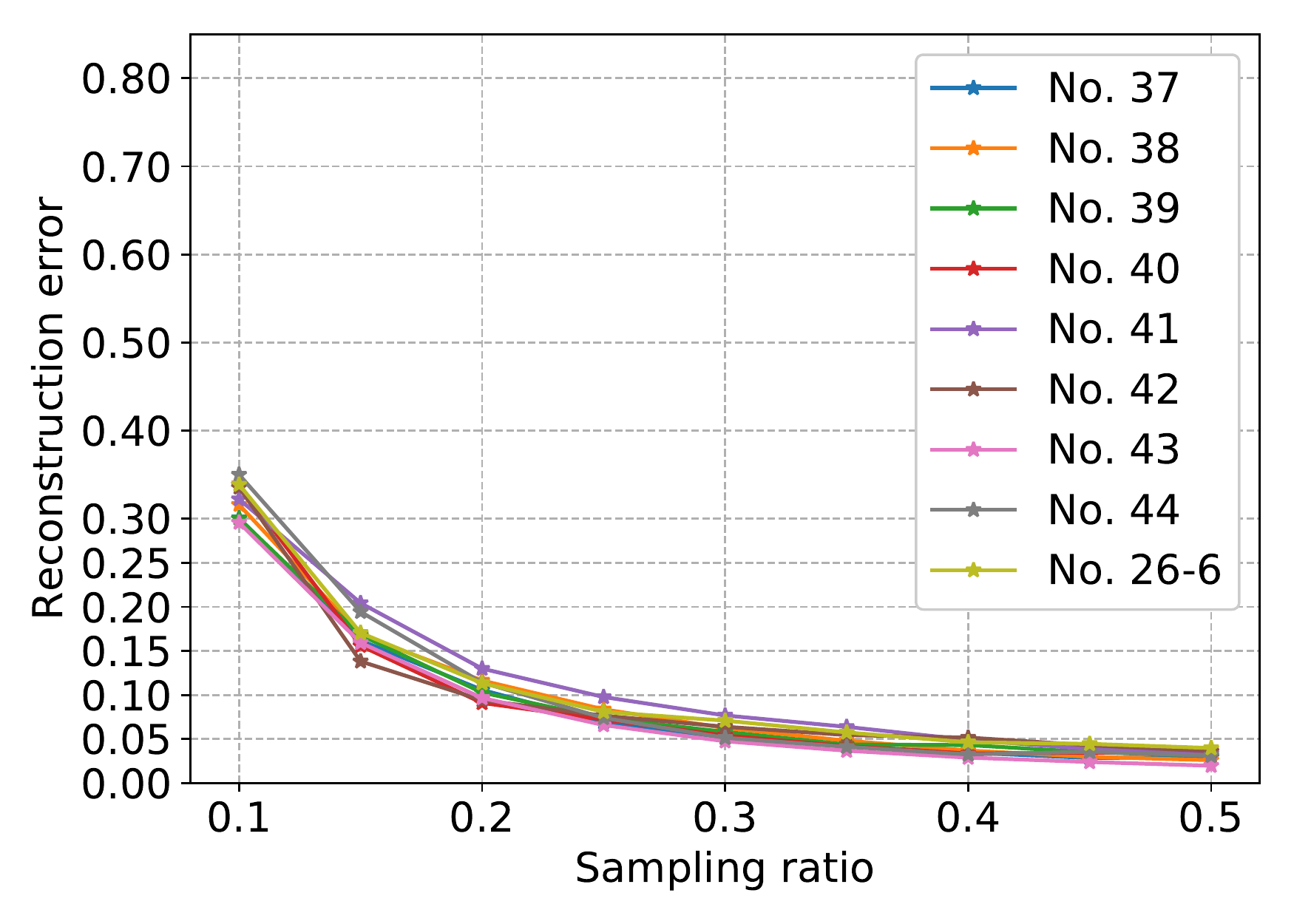}
\label{fig:9-3-2}
\end{subfigure}
\vspace{-1\baselineskip}
\end{figure}

\begin{figure}[H]
\setcounter{figure}{8}
\centering
\begin{subfigure}[b]{.38\linewidth}
\includegraphics[clip, trim=0cm 0.3cm 0cm 0.3cm, width=\linewidth]{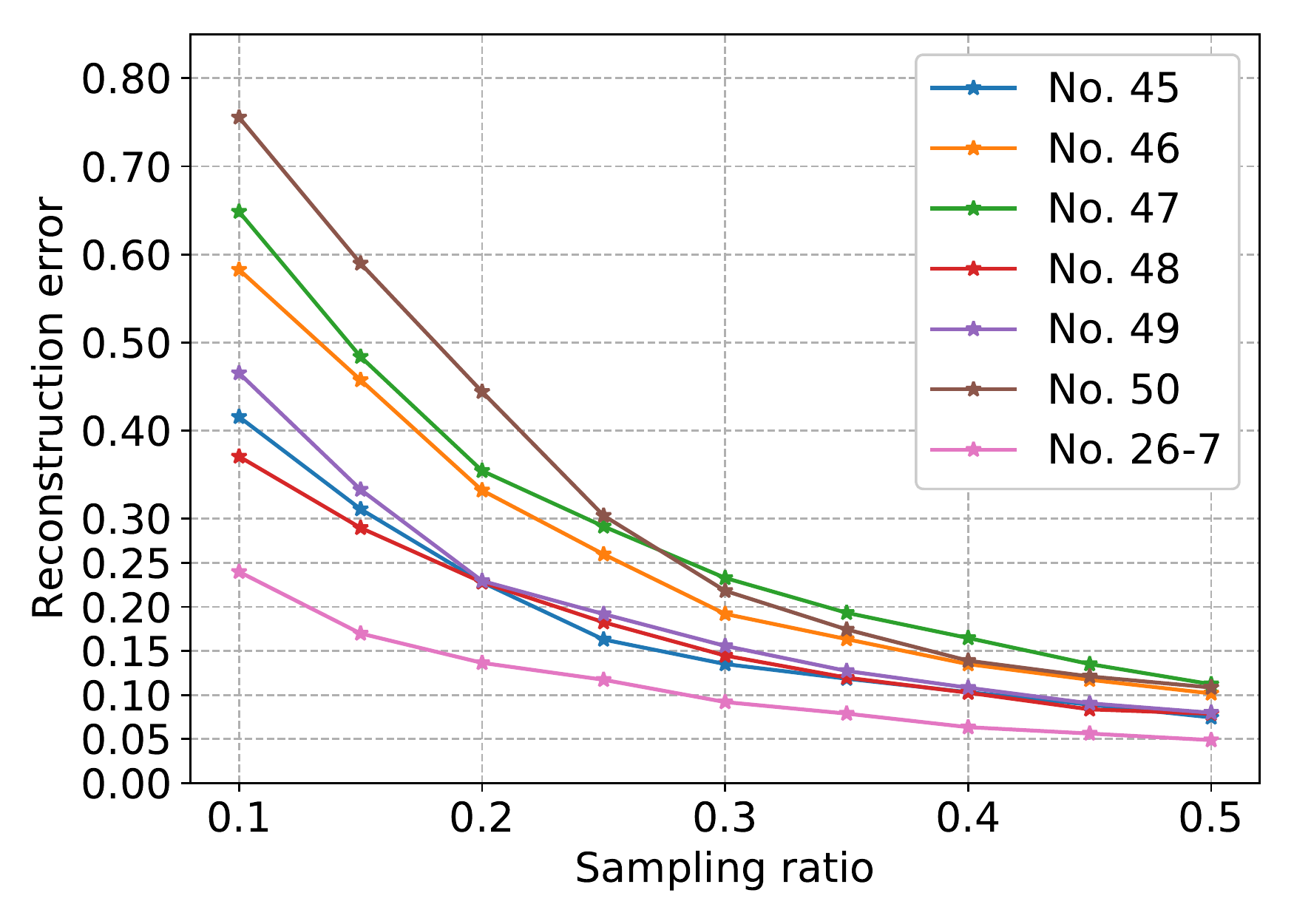}
\label{fig:9-4-1}
\end{subfigure}
\begin{subfigure}[b]{.38\linewidth}
\includegraphics[clip, trim=0cm 0.3cm 0cm 0.3cm, width=\linewidth]{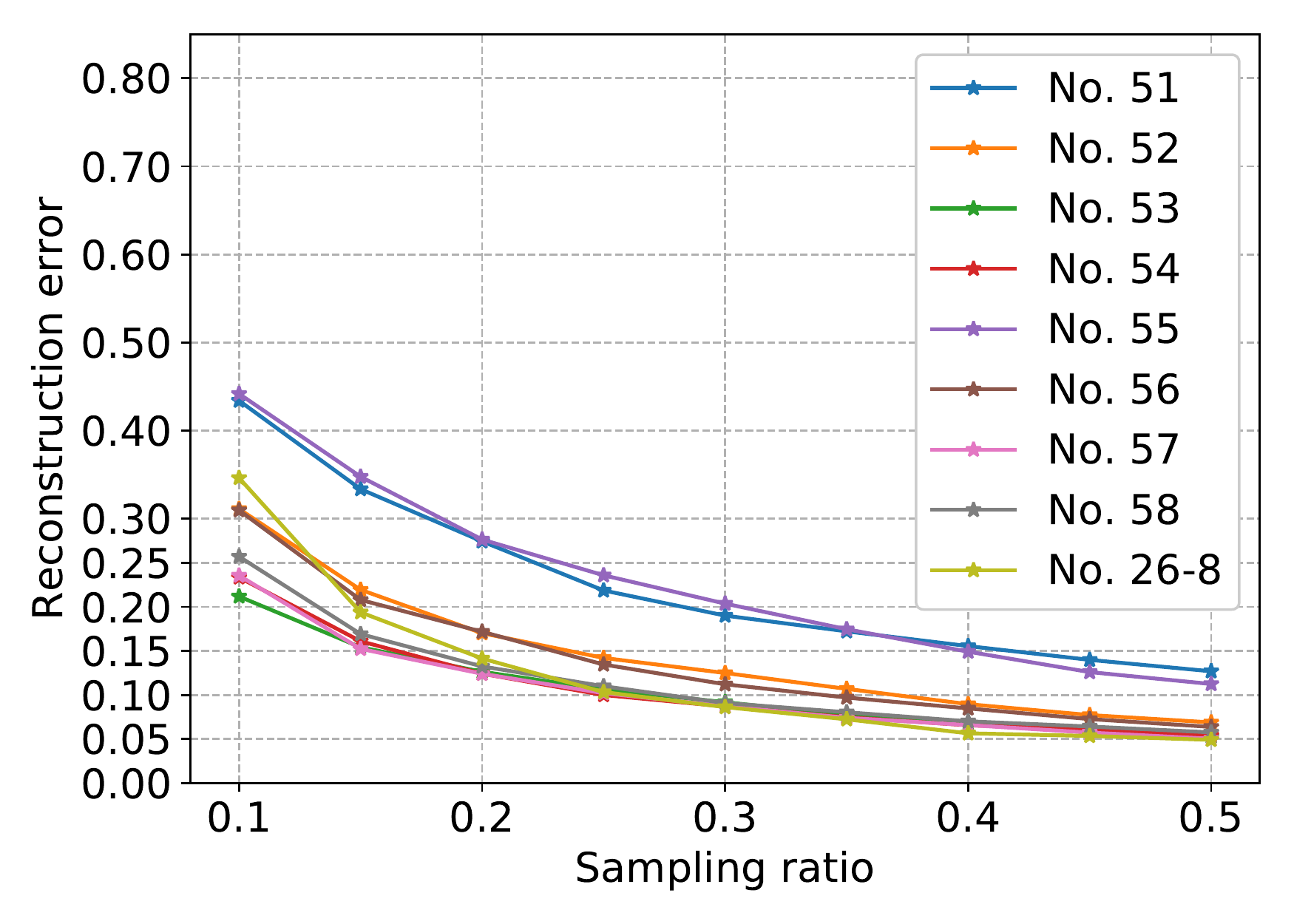}
\label{fig:9-4-2}
\end{subfigure}
\vspace{-1\baselineskip}

\begin{subfigure}[b]{.38\linewidth}
\includegraphics[clip, trim=0cm 0.3cm 0cm 0.3cm, width=\linewidth]{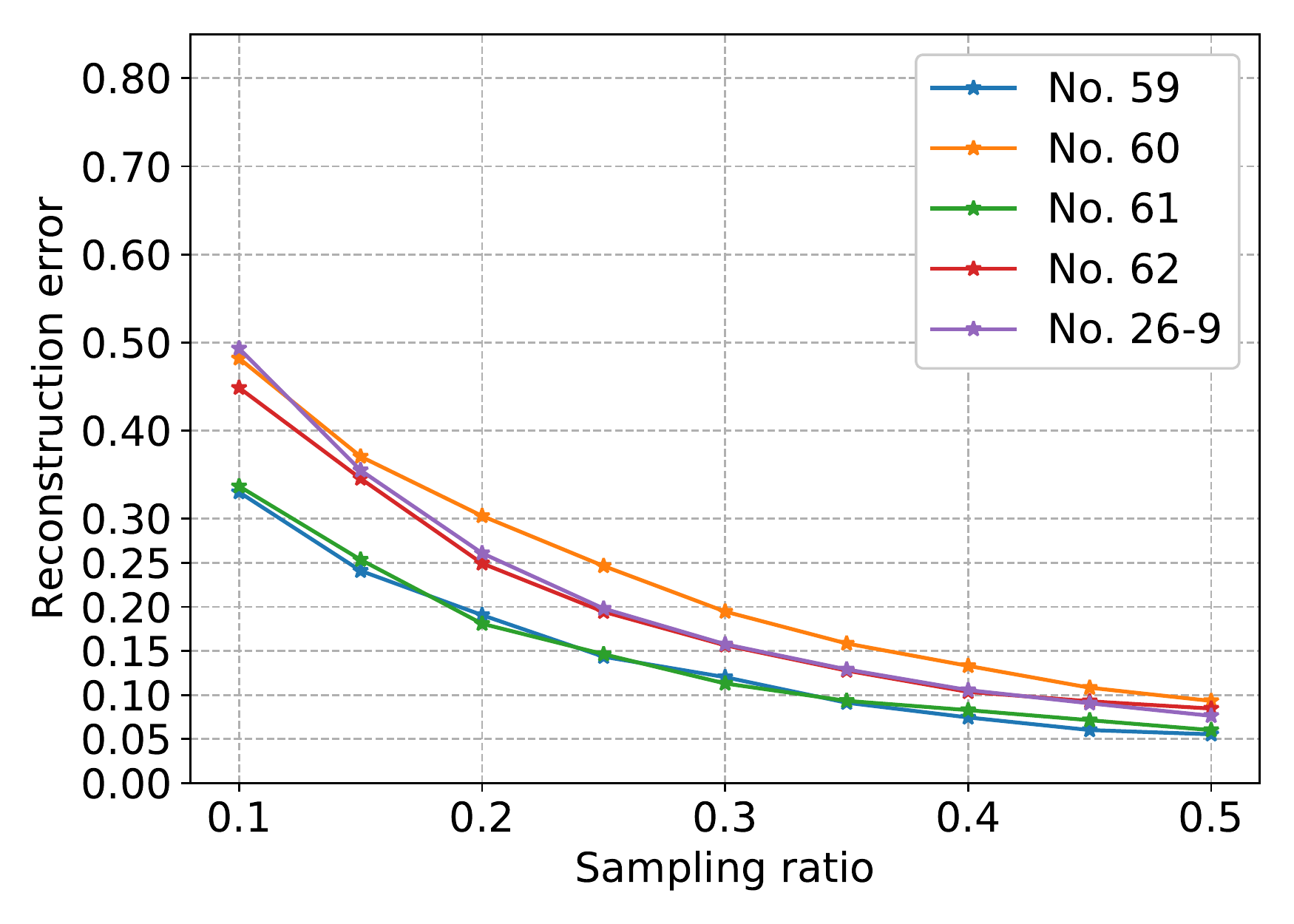}
\label{fig:9-5}
\end{subfigure}
\vspace{-1\baselineskip}
\caption{Simulation results based on field-test wireless data (nine tests are given by row)}
\label{fig:fig_9}
\end{figure}

\section{Discussion and conclusions}

In this paper, we proposed a novel neural network approach to solve the CS data-reconstruction problem. The prior knowledge, i.e., the Fourier-basis matrix and CS-sampled signals, are used as the input and the target of the network; the coefficient matrix is embedded as the parameters of a certain layer; and the objective function of conventional CS is set as the loss function of the network. Regularized by l1-norm, these basis coefficients are optimized to reduce the error between the original CS-sampled signals and masked reconstructed signals with a common gradient-descent optimization algorithm. Multiple signal channels can be reconstructed simultaneously. Validations using numerical data and field-test data show that data can be well reconstructed with low sampling ratio.

It is noteworthy that, although the neural network-based approach was used only to reconstruct the time-series SHM data, it is an open architecture for other conventional optimization problems. Essentially, a neural network can be considered as a computation graph, in which other objective functions or regularizations can be mapped into as elements such as nodes or weights. The integration of classical optimization problems and machine-learning techniques not only enables neural networks to be used to solve conventional optimization, but also makes the neural network interpretable and more trustworthy for users, which promotes the incorporation of machine-learning techniques into processes with well-grounded rationales or critical outputs. In the big data and AI times, machine learning will have a good potential to solve the optimization problems in civil engineering.

\section{Acknowledgment}

This research was supported by grants from the National Key R\&D Program of China (Grant No. 2017YFC1500603), the National Natural Science Foundation of China (Grant No. U1711265, 51678203 and 51638007).

\bibliographystyle{unsrt}
\bibliography{references}
\setcitestyle{number} 

\end{document}